\documentclass[12pt]{iopart}
\usepackage{graphics}
\usepackage{epsfig}
\usepackage{algorithmic}
\usepackage{iopams}  
\usepackage[usenames,dvipsnames]{color}

\usepackage{citesort}

\begin{document}

\title[Linearly scaling method for inverting banded matrices]{Linearly scaling direct method for accurately inverting sparse banded matrices}


\author{Pablo Garc\'ia-Risue\~no$^{1,2,3}$, Pablo Echenique$^{1,2,3,4}$}

\address{$^1$Instituto de Qu\'{\i}mica F\'{\i}sica Rocasolano, CSIC, Serrano 119, E-28006 Madrid, Spain}
\address{$^2$Departamento de F{\'{\i}}sica Te\'orica, Universidad de Zaragoza, Pedro Cerbuna 12, E-50009 Zaragoza, Spain}
\address{$^3$Instituto de Biocomputaci\'on y F{\'{\i}}sica de Sistemas Complejos (BIFI), Universidad de Zaragoza, Mariano Esquillor s/n, Edificio I+D, E-50018 Zaragoza, Spain}
\address{$^4$Unidad asociada IQFR-BIFI}

\ead{garcia.risueno@gmail.com} 

\begin{abstract}
In many problems in Computational Physics and Chemistry, one finds a special
kind of sparse matrices, called \emph{banded matrices}. These matrices, which
are defined as having non-zero entries only within a given distance from the
main diagonal, need often to be inverted in order to solve the associated
linear system of equations. In this work, we introduce a new $\mathcal{O}(n)$
algorithm for solving such a system, with the size of the
matrix being $n \times n$. We derive analytical recursive expressions that allow us to
directly obtain the solution. In addition,
we describe the extension to deal with matrices that
are banded plus a small number of non-zero entries outside the band, and we
use the same ideas to produce a method for obtaining the full inverse matrix.
Finally, we show that our new algorithm is competitive, both in accuracy and
in numerical efficiency, when compared with a standard method based on Gaussian
elimination. We do this using sets of large random banded matrices, as well as
the ones that appear in the calculation of Lagrange multipliers in proteins.

\end{abstract}

\vspace{2pc}
\submitto{\JPA}
\maketitle

\section{Introduction}
\label{sec_introduction}

In this article we present the efficient formulae and subsequent algorithms to
solve the system of linear equations
\begin{equation}
\label{basic_system}
Ax=b \ ,
\end{equation}
where $A$ is a $n \times n$ matrix, $x$ is the $n \times 1$ vector of 
unknowns, $b$ is a given $n \times 1$ vector and $A$ satisfies the equations below
for known values of
$m_{u}, m_{l}<n$
\begin{eqnarray}
A_{I, I+K} &=& 0 \quad \forall \ K>m_{u} \ , \quad \forall I  \ , \label{Aup}\\
A_{I+L,I} &=& 0 \quad \forall \ L>m_{l} \ , \quad \forall I  \ , \label{Adown}
\end{eqnarray}
i. e., $A$ is a \emph{banded matrix} and (\ref{basic_system}) is a
\emph{banded system}. We also investigate how to solve similar systems where
there are some non-zero entries not lying in the diagonal band.

Banded systems like this are abundant in computational physics and
computational chemistry literature, especially because the discretization of
differential equations, transforming them into finite-difference equations,
often results in banded matrices \cite{Don1987PC,Hym2002CG}. Many examples of
this can be found in boundary value problems
\cite{Sha1997IJCM,pap1991PC,Wri1992SIAM}, in fluid mechanics
\cite{Bri1977JCOP,Ari1992AM,Had2008}, thermodynamics \cite{Had2004E},
classical wave mechanics \cite{Sha1997IJCM}, structure mechanics
\cite{Pol2006PC}, nanoelectronics \cite{Pol2004JCOP}, circuit analysis
\cite{Lum1988}, diffusion equations and Maxwell's first-order curl equations
\cite{Hym2002CG}. In quantum chemistry, finite difference methods using banded
matrices are used both in wavefunction formalism
\cite{San1986JCOP,Gua1999JCP,Gua1999JCOP} and in density functional theory
\cite{Cas2006PSS,Mar2003CPC}. In addition to finite-difference problems,
banded systems 
are present in several
areas, such as constrained molecular simulation
\cite{Ryc1977JCOP,Alv2004JPCM,Alv2005JPCM}, including the calculation of
Lagrange multipliers in classical mechanics \cite{Maz2007JPA}. An important case
for the calculation of Lagrange multipliers deals with molecules with angular constraints.
The banded method presented here is suitable to
calculate the associated Lagrange multipliers exactly and efficiently \cite{GR2011JCC}. Banded matrix
techniques are useful not only in linear systems, but also in linearized ones,
which also appear frequently in the literature
\cite{pap1991PC,Bri1977JCOP,Ari1992AM,Had2008,Had2004E,Ryc1977JCOP}.

The solution of a linear system with $A$ being a $n \times n$ dense matrix
requires $\mathcal{O}(n^{3})$ floating point operations\footnote{As stated in \cite{Str1969NM}, this can be reduced to
$\mathcal{O}(n^{log_{2}7 \approx 2.807})$.} (a
floating point operation is an arithmetic operation, like addition, subtraction,
multiplication and ratio, involving real numbers which are represented in floating point notation, 
the customary nomenclature in computers). However, banded systems can
be solved in $\mathcal{O}(nm_{u}m_{l})$ floating point operations using very simple recursive
formulae, and the explicit form of $A^{-1}$ can be obtained in
$\mathcal{O}(n^{2})$ floating point operations. As mentioned earlier, there exist a number of
physical problems whose behaviour is described by banded systems where
$m_{u}, m_{l} \ll n$.
This makes it possible to get large computational savings if suitable algorithms are used,
what is even more important for computationally heavy problems
like those in which the calculation of relevant quantities
requires many iterations (Molecular Dynamics
\cite{Alo2008PRL,Gua1999JCP}, Monte Carlo simulations \cite{Has1970bio},
quantum properties calculations via self-consistent field equations
\cite{Ech2007MP,Gri1993PRA}, etc.).

In this work, we introduce a new algorithm for solving banded systems and
inverting banded matrices that presents very competitive numerical
properties, in many cases outperforming other commonly used techniques.
Additionally, we provide the explicit recursive expressions on which the
algorithm is based, thus facilitating further analytical developments.
The linear ($\mathcal{O}(n)$) scaling of the presented algorithm is a
remarkable feature since efficiency is commonly essential in today's
computer simulation of physical systems, specially in fields such as molecular
mechanics \cite{Pea1995CoPCo,claudioFF1,claudioFF2} (using efficient force
fields) and quantum ab initio methods
\cite{dft_linear_scaling_1,siesta2002,conquest_linear_dft}.

The article is structured as follows: in section \ref{sec_main}, we derive
simple recursive formulae for the efficient solution of a linear banded
system. These formulae enable the solution of (\ref{basic_system}) in
$\mathcal{O}(nm_{u}m_{l})$ floating point operations and are 
suitable to be used in a serial
machine. In section \ref{sec_sparse}, we extend these formulae to 
systems where some entries outside the band are also non-zero.
In section \ref{rad}, we briefly discuss the differences between our new algorithm and one
standard method to solve banded systems. In sec. \ref{secalltests} we quantitatively compare
the performance of both algorithms in terms of accuracy and numerical cost. 
For this comparison, in sec. \ref{numtests}
we use
randomly generated banded systems for inputs. In section \ref{sec_application} we
apply the algorithms to the problem of calculating the Lagrange multipliers which arise
when imposing holonomic constraints on proteins.
Finally, in section \ref{conclusions}, we state the most important
conclusions of the work. In the Appendix we provide
equations for the explicit expression of the entries of $A^{-1}$.
Some remarks on the algorithmic implementation of the methods presented here, their source code
and remarks on their parallelization can be found in the supplementary material.

\section{Analytical solution of banded systems}
\label{sec_main}

One of the most common ways of solving the linear system in
equation~(\ref{basic_system}) is by gradually changing the different entries of the 
matrix $A$ to zero through the procedure of \emph{Gaussian elimination}
\cite{And1999BOOK,Pre2007BOOK,WatXXXXBOOK}. This procedure is based on the
possibility of writing $A$ as $A=LU$, where $L$ is a lower triangular matrix and
$U$ is an upper triangular matrix. This way of writing $A$, called
$LU$-\emph{decomposition}, is possible (i.e., $L$ and $U$ exist), if, and only
if $A$ is invertible and all its leading principal minors are non-zero
\cite{Gol1989BOOK}. If one of the two matrices $L$ or $U$ is chosen to be
\emph{unit triangular}, i.e., with 1's on its diagonal, the matrices not only
exist but are also unique.

The analytical calculations and algorithms introduced in this work are based
on a different but closely related property of $A$, namely, the possibility of
finding $Q$ (a lower triangular matrix) and $P$ (an upper triangular one), so
that we have
\begin{equation}
\label{basis}
QAP=\mathbb{I} \quad \Rightarrow \quad A^{-1}= PQ \ ,
\end{equation} 
where $\mathbb{I}$ is the identity matrix.

The requirements in order for these two matrices to exist are the same as those in the
$LU$-decomposition, because in fact, the two propositions are equivalent.
The existence of a `$QP$-\emph{decomposition}' arises from the existence
of the $LU$ decomposition. This trivially proved if we make $Q=L^{-1}$ and $P=U^{-1}$.
The converse implication follows from the following facts. $A$
must be invertible so that equation (\ref{basic_system}) has a unique
solution. The fact that its determinant ($\det A$) is different from zero
and the relation $\det Q \det A \det P = \det\mathbb{I} = 1$ force both $Q$
and $P$ to have non-zero determinants and therefore to be invertible. This
enables one to write $A=Q^{-1}P^{-1}$, and since the inverse of a triangular matrix
is a triangular matrix of the same kind, we can identify $L=Q^{-1}$ and
$U=P^{-1}$ thus proving the existence of the $LU$-decomposition. This
equivalence also enables one to say that, as long as one of the two matrices $Q$
and $P$ is unit triangular, the $QP$-decomposition is unique.

An important qualification of this situation is that in order to solve the
system in equation~(\ref{basic_system}), using $QP$ (or $LU$) decomposition is not the only option.
We can also solve the system by performing a Gaussian
elimination process that is based on the $QP$ (or $LU$) decomposition of a
matrix $\tilde{A}$, which is obtained from $A$ by permuting its rows and/or
columns. If these permutations are performed (what is called \emph{pivoting}), the condition for
$Q\tilde{A}P=\mathbb{I}$ (or $\tilde{A}=LU$) to hold is simply that $A$ is
invertible.
Typically, the algorithms obtained from the pivoting case are 
more stable. For the sake of simplicity, derivations of this paper deal only with the non-pivoting case
(the reader should notice that pivoting can be included
in the debate with minor adjustments). 
In the supplementary material, algorithms including and lacking pivoting can be analysed.

Let us now build the matrices $P$ and $Q$ that satisfy
(\ref{basis}) for a given matrix $A$. When we have obtained them, they can be used to
compute the inverse $A^{-1}$, and then we will be able to solve
(\ref{basic_system}). However, in this section (see also
ref.~\cite{Cas2004JCP}) we will see that there is no need to explicitly build
$A^{-1}$, and the information needed to calculate $P$ and $Q$ can be used in a
different way to solve (\ref{basic_system}).

We begin by writing $P$ and $Q$ as follows
\numparts
\begin{eqnarray}\label{PQcompletas}
P & := P_{1}P_{2} \ldots P_{n}=\prod_{K=1}^{n}P_{K} \ , \label{Pcompleta}\\
Q & := Q_{n}Q_{n-1}\ldots Q_{1}=\prod_{K=n}^{1}Q_{K} \ , \label{Qcompleta}
\end{eqnarray} 
\endnumparts
being
\begin{equation}
\label{defP}
P_{K}:=
\left(\begin{array}{ccccccccc}
1 & & & & & & & & \\
  & \ddots & & & & & & & \\
  & & 1 & & & & & & \\
  & & & \xi_{KK} & \xi_{K,K+1} & \ldots & \xi_{K,K+m_{u}} & & \\
  & & & & 1 & & & & \\
  & & & & & \ddots & & & \\
  & & & & & & 1 & & \\
  & & & & & & & \ddots & \\
  & & & & & & & & 1
\end{array} \right) \ ,
\end{equation}
and
\begin{equation}
\label{defQ}
Q_{K}:=
\left(\begin{array}{cccccccc}
1 & & & & & & & \\
  & \ddots & & & & & & \\
  & & 1 & & & & & \\
  & & \xi_{K+1,K} & 1 & & & & \\
  & & \vdots & & \ddots & & & \\
  & & \xi_{K+m_{l},K} & & & 1 & & \\
  & & & & & & \ddots & \\
  & & & & & & & 1
\end{array} \right) \ ,
\end{equation}
where $K=1,\ldots,n$, and all the non-specified entries are zero. Note that 
$P_{K}$ equals the identity matrix except in its $K$-th row, and
$Q_{K}$ equals the identity matrix except in its $K$-th column.

Now, the trick is to choose all coefficients $\xi_{IJ}$ in the preceding
matrices so that we have $QAP=\mathbb{I}$ in (\ref{basis}) (whenever the
conditions for this to be possible are satisfied; see the beginning of this
section). 

First we must notice that given (\ref{defP}), multiplying a generic
matrix $G$ (by its right) by $P_{K}$ is equivalent to adding the $K$-th column
of $G$ multiplied by the corresponding $\xi$ coefficients to several columns
of $G$, while at the same time multiplying the $K$-th column of the
original matrix by $\xi_{KK}$: 
\numparts
\begin{eqnarray}
\label{GP}
(GP_{K})_{IJ} & = G_{IJ}  \ & \qquad \textrm{for $J < K$ and $J > K+m_{u}$} \ ,
 \label{GP_a} \\ 
(GP_{K})_{IK} & = G_{IK}\xi_{KK} \ , & \qquad \label{GP_b}\\
(GP_{K})_{IJ} & = G_{IJ}+G_{IK}\xi_{KJ}  \  & \qquad \textrm{for } K<J \leq K+m_{u}
 \ . \label{GP_c}
\end{eqnarray}
\endnumparts
If we take this into account, we can choose $\xi_{11}$ so that
$(AP_{1})_{11}=1$, and $(AP_{1})_{1J}=0$ for $J=2, \ldots, n$. Given the fact
that $A$ is banded (see, in particular (\ref{Aup}, \ref{Adown})), we have
that
\numparts
\begin{eqnarray}
\label{ap1}
 (A P_1)_{11}  = A_{11}\xi_{11}=1 \ &\Rightarrow \ \xi_{11}=1/A_{11} \ , \\
 (A P_1)_{1J}  = A_{1J}+A_{11}\xi_{1J}=0 \ &\Rightarrow \  
 \xi_{1J}=-\frac{A_{1J}}{A_{11}} \ , \quad
 1 < J \leq 1+m_{u} \ .
\end{eqnarray}
\endnumparts
Operating in this way, we have `erased' (i.e., turned into zeros) the
superdiagonal entries\footnote{{ We call \emph{superdiagonal} entries of a matrix 
$A$ to the entries $A_{IJ}$ with $I<J$, and \emph{subdiagonal} entries to the entries 
$A_{IJ}$ with $I>J$.} } 
of $A$ that lie on its first row, and we have done this
by multiplying $A$ on the right by $P_{1}$ with the appropriate $\xi_{1J}$.
Then, if we multiply $(AP_1)$ on the right by $P_2$ and choose the coefficients
$\xi_{2J}$ in the analogous way, we can erase all the superdiagonal entries in
the second row and change its diagonal entry to 1. In general, multiplying 
${(A P_1 \cdots P_{K-1})}$ by $P_K$ erases the superdiagonal entries of the $K$-th
row of ${(A P_1 \cdots P_{K-1})}$, and turns its diagonal $KK$ entry to 1.
This procedure is called
\emph{Gaussian elimination} \cite{Gol1989BOOK}, and after $n$ steps, the
resulting matrix is a lower unit triangular matrix
$A\prod_{K=1}^{n}P_{K}=AP$.

The expression for the coefficients $\xi_{IJ}$, with $I \leq J$ and $I > 1$ is
more complex than (\ref{ap1}) because, as a consequence of (\ref{GP_a}, \ref{GP_b}, \ref{GP_c}),
whenever we multiply a matrix on the right by $P_{K}$, not only is its $K$-th row
(the one we are erasing) affected, but also all the rows below are affected (the $m_{l}$
rows below in the case of a banded matrix like (\ref{basis})). However, the
matrix $A\prod_{L=1}^{K-1}P_{L}$ is 0 in all its superdiagonal entries
belonging to the first $K-1$ rows, and multiplying it on the right by $P_{K}$
has no influence on these rows. Hence, the fact that we have chosen to erase
the superdiagonal entries of $A$ from the first row to the last row allows us
to express the general conditions that the $\xi$ coefficients belonging to
different $P_{K}$'s must satisfy the following:
\numparts
\begin{eqnarray}
\label{defeps}
\left(A \prod_{K=1}^{I}P_{K}\right)_{II} &=1 \ , 
 \label{defeps_a} \\
\left(A \prod_{K=1}^{I}P_{K}\right)_{IJ} &=0 \quad \textrm{for $I<J$} \ . 
 \label{defeps_b}
\end{eqnarray}
\endnumparts
Now, using (\ref{defeps_a}) together with (\ref{GP_b}), we can derive
the following expression for the coefficient $\xi_{II}$ in terms of the
previous steps of the process:
\begin{eqnarray}
\label{jj1}
\left(A \prod_{K=1}^{I}P_{K}\right)_{II} & = &
\left(A \prod_{K=1}^{I-1}P_{K}P_{I}\right)_{II} =
\left(A \prod_{K=1}^{I-1}P_{K}\right)_{II} \xi_{II} = 1
 \nonumber \\
 &  & \Longrightarrow \quad
 \xi_{II} = \frac{1}{\left(A \prod_{K=1}^{I-1}P_{K}\right)_{II}} \ .
\end{eqnarray}

Analogously, using (\ref{defeps_b}) and (\ref{GP_c}), we can write an
explicit expression for $\xi_{IJ}$ with $I < J$:
\begin{eqnarray}
\label{jj2}
\left(A \prod_{K=1}^{I-1}P_{K}P_{I}\right)_{IJ} & = &
\left(A \prod_{K=1}^{I-1}P_{K}\right)_{IJ} +
\left(A \prod_{K=1}^{I-1}P_{K}\right)_{II}\xi_{IJ} = 0
\nonumber \\
 \Longrightarrow \quad
\xi_{IJ} & = & -\frac{\left(A \prod_{K=1}^{I-1}P_{K}\right)_{IJ}}
                 {\left(A \prod_{K=1}^{I-1}P_{K}\right)_{II}}
 = - \left(A \prod_{K=1}^{I-1}P_{K}\right)_{IJ}\xi_{II} \ .
\end{eqnarray}

Also according to (\ref{GP_a}, \ref{GP_b}, \ref{GP_c}), for $I \leq J$
\begin{equation}\label{jjj}
\left(A\prod_{K=1}^{L}P_{K}\right)_{IJ}  = A_{IJ}+
\sum_{M=J-m_{u}}^{L}\left(A\prod_{K=1}^{M-1}P_{K}\right)_{IM}\xi_{MJ} \ ,
 \quad I > L \ .
\end{equation}

Note that, in this equation we have $I > M$\footnote{Because $I>L$ by hypothesis, and $L \geq M$.},
which entails that $\left(A\prod_{K=1}^{M-1}P_{K}\right)_{IM}$ are subdiagonal entries.
This enables one to calculate
the coefficients $\xi_{IJ}$ with $I > J$, i.e., those that correspond to the
matrices $Q_{K}$, once all the coefficients in the matrices $P_{K}$ have
already been evaluated. We know that $AP$ is a unit lower triangular matrix.
This means that its subdiagonal $I,M$ entry (with $I > M$) equals $\xi_{IJ}$,
because no other changes affect this entry when multiplying $AP$ by the
different $Q_{K}$'s. 
{
If $G$ is a generic matrix, then (\ref{defQ}) implies 
\numparts
\begin{eqnarray}\label{QG}
(Q_{K}G)_{IJ} & = G_{IJ}  \ & \qquad  \textrm{for $I \leq K$ and $I > K+m_{l}$} \ ,  \label{QG_a} \\ 
(Q_{K}G)_{IJ} & = G_{IJ}+G_{KJ}\xi_{IK}  \  &  \qquad \textrm{for $K < I \leq K+m_{l}$, for all $J$} \ . \label{QG_b}
\end{eqnarray}
\endnumparts
If $T$ is any unit lower triangular matrix (this is, its 
diagonal entries equal 1 and its superdiagonal entries are zero, $T_{II}=1$, $T_{IJ}=0$ for $I<J$),
then (\ref{defQ}) implies 
\numparts
\begin{eqnarray}\label{QT}
(Q_{K}T)_{IJ} & = T_{IJ}  \ & \qquad  \textrm{for $I \leq K$ and $I > K+m_{l}$} \ ,  \label{QT_a} \\ 
(Q_{K}T)_{IJ} & = T_{IJ}+T_{KJ}\xi_{IK}  \  &  \qquad \textrm{for $K < I \leq K+m_{l}$, $J\leq K$} \ , \label{QT_b} \\
(Q_{K}T)_{IJ} & = T_{IJ}  \  &  \qquad \textrm{for $K < I \leq K+m_{l}$, $J > K$} \ . \label{QT_c}
\end{eqnarray}
\endnumparts
Moreover, if $S_{K-1}$ is a unit lower triangular matrix satisfying $(S_{K-1})_{IJ}=0$ for $I>J$, $J<K$, then
equations  (\ref{QT_a}, \ref{QT_b}, \ref{QT_c}) become
\numparts
\begin{eqnarray}\label{ST}
(Q_{K}(S_{K-1}))_{IJ}  = & (S_{K-1})_{IJ}  \  \qquad  \textrm{for $I \leq K$ and $I > K+m_{l}$} \ ,  \label{QS_a} \\ 
(Q_{K}(S_{K-1}))_{IJ}  = & (S_{K-1})_{IK}+(S_{K-1})_{KK}\xi_{IK}  \    \ \textrm{for $K < I \leq K+m_{l}$} \ , \label{QS_b} \\
(Q_{K}(S_{K-1}))_{IJ}  = & (S_{K-1})_{IJ}  \    \qquad \textrm{for $K < I \leq K+m_{l}$, $J > K$} \ . \label{QS_c}
\end{eqnarray}
\endnumparts
Multiplying $Q_1$ on the left by $(AP)$ erases the subdiagonal entries of the first column of $(AP)$, and
multiplying $(Q_1AP)$ on the left by $Q_2$ erases the subdiagonal entries of the second column of $(Q_1AP)$.
By repeating this procedure,
multiplying $Q_K$ by $(Q_{K-1} \cdots Q_1 AP)$ by the left, the subdiagonal entries of the $K$ column of
$(Q_{K-1} \cdots Q_1 AP)$ are erased. Therefore, $(Q_{K-1} \cdots Q_1 AP)$ satisfies the conditions of $S_{K-1}$,
and hence it satisfies equations (\ref{QS_a}, \ref{QS_b}, \ref{QS_c}). The conditions for the correct erasing
are:
\begin{eqnarray}\label{SST}
& (Q_{K} (Q_{K-1} \cdots Q_1 AP) )_{IK}   =  \label{SST_a}\\ 
& (Q_{K-1} \cdots Q_1 AP)_{IK}+(Q_{K-1} \cdots Q_1 AP)_{KK}\xi_{IK} = 0 \    \ \textrm{for $K < I \leq K+m_{l}$} \ .  \nonumber
\end{eqnarray}
The expressions in (\ref{SST_a}) can be simplified. Equation (\ref{QS_c}) implies
\numparts
\begin{eqnarray}\label{SST}
& (Q_{K-1} \cdots Q_1 AP)_{IK} = (Q_{K-1} (Q_{K-2}\cdots Q_1 AP) )_{IK} =  & \nonumber \\ 
& (Q_{K'} (Q_{K'-1}\cdots Q_1 AP) )_{I,K'+1} = (Q_{K'-1}\cdots Q_1 AP)_{I,K'+1} =  &  \nonumber \\ 
& (Q_{K-2}\cdots Q_1 AP)_{IK} \ , &
\end{eqnarray}
\endnumparts
where we have defined $K':=K-1$. By repeating operations like this, it is easy to obtain
\numparts
\begin{eqnarray}
(Q_{K-1} \cdots Q_1 AP)_{IK} = (AP)_{IK}  \   \qquad \textrm{for $K < I \leq K+m_{l}$} \ ,  \label{cuca1}\\
(Q_{K-1} \cdots Q_1 AP)_{KK} = (AP)_{KK} = 1 \ .  \label{cuca2}
\end{eqnarray}
\endnumparts
In addition, we must consider that (\ref{GP_a}, \ref{GP_b}, \ref{GP_c}) imply
\begin{equation}\label{finaldeq}
\left(A\prod_{L=1}^{M-1}P_{L}\right)_{IM} = \left(A\prod_{L=1}^{n}P_{L}\right)_{IM} / \xi_{MM} = (AP)_{IM} / \xi_{MM} \qquad \textrm{for $I>M$} \ .
\end{equation}
Using (\ref{cuca1}, \ref{cuca2}, \ref{finaldeq}) into (\ref{SST_a}) we obtain
} 
%
%
%
\begin{equation}
\left(A\prod_{L=1}^{M-1}P_{L}\right)_{IM}=-\xi_{IM} / \xi_{MM} \quad
\textrm{ for } I > M \ .
\label{subd}
\end{equation}

If we apply this on the right hand side of (\ref{jjj}), then insert the
resulting expression with $J=I$ and $L=I-1$ into (\ref{jj1}) and also insert
it with $L=I-1$ into (\ref{jj2}), we get the following recursive equations
\numparts
\begin{eqnarray}\label{cccc}
\xi_{II} & = \left(A_{II} - 
  \sum_{M=I-m_{u}}^{I-1}\frac{\xi_{IM}}{\xi_{MM}}\xi_{MI} \right)^{-1} \ , \label{xiarriba2c} \\
\xi_{IJ} & = \xi_{II}\left(-A_{IJ} + 
  \sum_{M=I-m_{u}}^{I-1}\frac{\xi_{IM}}{\xi_{MM}}\xi_{MJ}\right) \quad
  \textrm{for } I < J \ . \label{xiarriba1c}
\end{eqnarray}
\endnumparts
Following an analogue procedure, we obtain
\begin{equation}
\xi_{IJ} = \xi_{II}\left(-A_{IJ} + \sum_{M=J-m_{l}}^{J-1}\frac{\xi_{IM}}{\xi_{MM}}\xi_{MJ}\right) \quad
  \textrm{for } I > J \ . \label{xiarriba1ccc}
\end{equation}
For efficiency reasons, we prefer to define $\chi_{IJ}:=\xi_{IJ}/\xi_{II}$ for $I>J$. This makes 
(\ref{xiarriba2c}, \ref{xiarriba1c}, \ref{xiarriba1ccc}) become
\numparts
\begin{eqnarray}\label{ccc}
\xi_{II} & = \left(A_{II} - 
  \sum_{M=I-m_{u}}^{I-1}\chi_{IM}\xi_{MI} \right)^{-1} \ , 
  \label{ccc1} \\
\xi_{IJ} & = \xi_{II}\left(-A_{IJ} + 
  \sum_{M=I-m_{u}}^{I-1}\chi_{IM}\xi_{MJ}\right) \quad
  \textrm{for } I<J \ , \label{ccc2} \\
\chi_{IJ} & = -A_{IJ} + 
  \sum_{M=J-m_{l}}^{J-1}\chi_{IM}\xi_{MJ} \quad
  \textrm{for } I>J \ . \label{ccc3}
\end{eqnarray}
\endnumparts


The three equations above can be further modified with the aim of improving the numerical
efficiency of the algorithms derived from them. The starting point for the 
summations in (\ref{ccc}) must be the value of $M$ such that both $\xi_{IM}$ 
and $\xi_{MJ}$ are non-zero. We must take into account that in a banded matrix 
the number of non zero entries above and on the left of the $I,J$ entry 
depends on the values of $I$, $J$:
\begin{itemize}
\item There are $m_{u}+(I-J)$ non-zero entries immediately above $A_{IJ}$.
\item There are $m_{l}-(I-J)$ non-zero entries immediately on the left of 
      $A_{IJ}$.
\end{itemize}

These properties are also satisfied in $A\prod_{L=1}^{K}P_{L}$ for all $K$. 
Therefore, if we define
\begin{eqnarray}
\mu_{IJ} & := & \min\{m_{u}+(I-J), m_{l}-(I-J)\} \ , \nonumber \\
\mu^\prime & := & \min\{m_{u}, m_{l}\} \ , \nonumber
\end{eqnarray}
we can re-express (\ref{ccc1}, \ref{ccc2}, \ref{ccc3}) as
\numparts
\begin{eqnarray}\label{xicoefsfinal}
\xi_{II} & = \left( A_{II} - 
 \sum_{M=\max\{1,I-\mu^\prime\}}^{I-1}\chi_{IM}\xi_{MI} \right)^{-1} \ , 
 \label{xifinal_a} \\
\xi_{IJ} & = \xi_{II}\left(-A_{IJ} + 
 \sum_{M=\max\{1,I-\mu_{IJ}\}}^{I-1} \chi_{IM}\xi_{MJ}\right)
 & \quad \textrm{for } I < J \ , \label{xifinal_b} \\
\chi_{IJ} & = -A_{IJ} + 
 \sum_{M=\max\{1,J-\mu_{IJ}\}}^{J-1}\chi_{IM}\xi_{MJ}
 &  \quad \textrm{for } I > J \ . \label{xifinal_c}
\end{eqnarray}
\endnumparts

In the restricted but very common case in which $m_{l}=m_{u}=:m$, the previous 
equations become
\numparts
\begin{eqnarray}
\xi_{II} & = \left(A_{II} -
 \sum_{M=max(1,I-m)}^{I-1}\chi_{IM}\xi_{MI} \right)^{-1}
 \ , \label{xifinal_a2} \\
\xi_{IJ} & = \xi_{II}\left(-A_{IJ} + 
 \sum_{M=\max\{1,J-m\}}^{I-1}\chi_{IM}\xi_{MJ}\right)
 & \textrm{for } I < J \ , \label{xifinal_b2} \\
\chi_{IJ} & = -A_{IJ} +
 \sum_{M=\max\{1,I-m\}}^{J-1}\chi_{IM}\xi_{MJ} 
 & \textrm{for } I > J \ . \label{xifinal_c2}
\end{eqnarray}
\endnumparts

If the matrix $A$ is symmetric ($A_{IJ}=A_{JI}$), we can avoid performing many operations simply by using
\begin{equation}
\chi_{IJ}=\xi_{JI}/\xi_{JJ} \ , \qquad \textrm{for} \quad I > J
 \ , \label{xifinal_csim} 
\end{equation}
instead of (\ref{xifinal_c}). Equation (\ref{xifinal_csim}) can easily 
be obtained from (\ref{xicoefsfinal}) by induction. 

The reader must also note that, although the coefficients $\xi_{IJ}$ have been
obtained by performing the products $\prod_{K=n}^{1}Q_{K}A\prod_{L=1}^{n}P_{L}$
in a certain order, they are independent of this choice. Indeed, if we take
a look to expressions (\ref{Pcompleta}), (\ref{Qcompleta}), (\ref{defP}), and
(\ref{defQ}), we can see that the $K$-th row (or column) is always erased
before the $(K+1)$-th one. 
It does not matter if we apply first $Q_{K}$ or
$P_{K}$ to erase the $K$ row (or column); the result of the
operation will be the same. In both cases, $-G_{IK}G_{KJ}/G_{KK}$ (where
$G:=\prod_{M=K-1}^{1}Q_{M}A\prod_{L=1}^{K-1}P_{L}$) is added to all the entries of
$G_{IJ}$ such that $I \in \{K+1, \ldots, K+m_{l}\}$ and $J \in \{K+1, \ldots,
K+m_{u}\}$. This is valid when both the $K$-th row and the $K$-th column are
not erased yet. If one of them is already erased, erasing the other has no
influence on $G_{IJ}$ with $I \in \{K+1, \ldots, K+m_{l}\}$ and $J \in \{K+1,
\ldots, K+m_{u}\}$. In both cases $\xi_{IK}=-G_{IK}/G_{KK}$ for $I>K$, and
$\xi_{KJ}=-G_{KJ}/G_{KK}$ for $J>K$. This is because all the previous rows (or
columns) have been nullified before, and adding columns (or rows) has no influence
on the $K$-th one.

Now, the algorithm to solve (\ref{basic_system}) can be divided into three 
stages (in our implementation we join together the first and second ones). 
Since $A^{-1}=PQ$ (\ref{basis}), these steps are:

\begin{enumerate}

\item To obtain the coefficients $\xi$.

\item To obtain the intermediate vector $c:=Qb$.

\item To obtain the final vector $x=Pc$.

\end{enumerate}

Now, using the results derived above, let us calculate the expressions for the 
second and third steps:

Whenever we multiply a generic $n \times 1$ vector $v$ on the left by $Q_{K}$
(see (\ref{defQ})), we modify its $K$-th to $(K+m_{l})$-th rows in the
following way:
\numparts
\begin{eqnarray}\label{6MB}
(Q_{K}v)_{I} & = v_{I} & \qquad  \textrm{for } I \leq K,\quad I > K+m_{l}
  \ ,\label{6MB1} \\
(Q_{K}v)_{I} & = v_{I} + \xi_{IK}v_{K} & \qquad \textrm{for } K<I \leq K+m_{l}
  \ . \label{6MB2}
\end{eqnarray}
\endnumparts
Since $Q:=Q_{n}Q_{n-1}\ldots Q_{1}$, using the expression for each of the
$Q_K$ in (\ref{defQ}), and the fact that $\xi_{IJ}=0$ for $I>J+m_{l}$, we have
\numparts
\begin{eqnarray}
Q_{IJ} & = 0 & \qquad \textrm{for } I<J \ , \label{q0} \\
Q_{II} & = 1 \ , & \qquad  \label{q1} \\
Q_{IJ} & = \sum_{M=\max\{I-m_{l},1\}}^{I-1}\xi_{IM}Q_{MJ}
  & \qquad  \textrm{for } I>J \ . \label{qij}
\end{eqnarray}
\endnumparts
where the maximum in the lower limit of the sum accounts for boundary effects
and ensures that $M$ is never smaller than $1$.

From these relations between the entries of $Q$, we can get the components 
$c_{I}$ of the intermediate vector $c$ in the second step above: 
\begin{eqnarray}\label{cfinal}
c_{I} & = & \sum_{J=1}^{n}Q_{IJ}b_{J} = \sum_{J=1}^{I}Q_{IJ}b_{J} = 
b_{I} + \sum_{J=1}^{I-1}\left(\sum_{M=\max\{I-m_{l},1\}}^{I-1}
 \xi_{IM}Q_{MJ}\right)b_{J} \nonumber \\
 & = & b_{I} + \sum_{M=\max\{I-m_{l},1\}}^{I-1} \xi_{IM} 
 \sum_{J=1}^{I-1}Q_{MJ}b_{J} =  b_{I} + \sum_{M=\max\{I-m_{l},1\}}^{I-1}\xi_{IM}c_{M}  \nonumber \\
 & = &
 b_{I} + \sum_{M=\max\{I-m_{l},1\}}^{I-1}\chi_{IM}\xi_{MM}c_{M} \ . \label{ci}
\end{eqnarray}
In the first row of the equation above, we applied (\ref{q0}), and then (\ref{qij}).
In the second row of the equation above, we performed a feedback in the equation.

We will now turn to the third and final step of the process, which consists 
of calculating the final vector $x=Pc$.
Whenever we multiply a generic $n \times n$ matrix $G$ on the left by $P_{K}$
(see equation~(\ref{defP})), the resulting matrix is the same as $G$ in all its
rows except for the $K$-th one, which is equal to a linear combination of the
first $m_u+1$ rows below it:
\numparts
\begin{eqnarray}\label{PGMB}
(P_{K}G)_{IJ} & = G_{IJ} & \textrm{for $I\neq K$} \ , \\
(P_{K}G)_{KJ} & = \sum_{L=K}^{\min\{K+m_u,n\}}\xi_{KL} G_{LJ} \ , &
\end{eqnarray}
\endnumparts
where the minimum in the upper limit of the sum accounts for boundary effects
and ensures that $K+L$ is never larger than $n$.

If we now use the relations above to construct $P$ as in (\ref{Pcompleta}),
i.e., we first take $P_n$ and multiply it on the left by $P_{n-1}$, then we
multiply the result, $P_{n-1}P_n$, on the left by $P_{n-2}$, etc., we
arrive to:
\numparts
\begin{eqnarray}\label{4MB}
P_{II} & =\xi_{II} \ , & \qquad \\
P_{IJ} & =\sum_{K=I+1}^{\min\{I+m_{u},n\}}\xi_{IK}P_{KJ} & \qquad \textrm{for } I<J
 \ , \label{pij} \\
P_{IJ}&=0 & \qquad \textrm{for } I>J \ ,
\end{eqnarray}
\endnumparts
meaning that every row of $P$ is a linear combination of the following rows,
plus a term in the diagonal.

These expressions allow us to obtain the last equation that is needed to 
solve the linear system in (\ref{basic_system}):
\begin{eqnarray}
\label{xfinal}
x_{I} & = & \sum_{J=1}^{n}P_{IJ}c_{J} = \sum_{J=I}^{n}P_{IJ}c_{J} 
 = \xi_{II}c_{I} + \sum_{J=I+1}^{n} 
   \left(\sum_{K=I+1}^{\min\{I+m_{u},n\}} \xi_{IK}P_{KJ}\right)c_{J}
   \nonumber \\
 & = & \xi_{II}c_{I} + 
 \sum_{K=I+1}^{\min\{I+m_{u},n\}} \xi_{IK}
 \sum_{J=I+1}^{n}P_{KJ}c_{J} \nonumber \\
 & = & \xi_{II}c_{I}+\sum_{K=I+1}^{\min\{I+m_{u},n\}}\xi_{IK}x_{K}
  \label{xi} \ .
\end{eqnarray}

Now, we can use expressions (\ref{xifinal_a}), (\ref{xifinal_b}), and
(\ref{xifinal_c}) in order to obtain the coefficients $\xi$, and then plug
them into (\ref{ci}) and (\ref{xi}) in order to finally solve (\ref{basic_system}).

To conclude, let us focus on the computational cost of this
procedure. From (\ref{xifinal_a}), (\ref{xifinal_b}), and (\ref{xifinal_c}),
it follows that obtaining the coefficients $\xi$ requires $\mathcal{O}(n)$
floating point operations. Being more precise, the summations in (\ref{xifinal_a}),
(\ref{xifinal_b}), and (\ref{xifinal_c}), require $\mu_{IJ}$ products and
$\mu_{IJ}-1$ additions ($2\mu_{IJ}-1$ floating point operations)\footnote{The meaning of 
$m_u, m_l$ can be noticed in (\ref{Aup}, \ref{Adown}).}. If, without loss of generality,
we consider $m_{u} \geq m_{l}$, it is easy to check that the following
computational costs hold:

\begin{itemize}

\item Obtaining one diagonal $\xi_{II}$ takes $2 \mu' + 2 \simeq 2 m_{u}$ 
floating point operations.

\item Obtaining one superdiagonal coefficient $\xi_{IJ}$ (where $I<J$) takes
about $2\min\{m_{l},m_{u}-(J-I)\}$ floating point operations. Hence, in order to obtain all the
coefficients in a column above the diagonal, there are two sets of $\xi$'s
that require a different number of operations. The lower one requires $2m_{l}$
floating point operations and the upper one requires $2(m_{u}-(J-I))$ floating point operations. All in all, obtaining
$\xi_{J-I,J}$ for $I=1, \ldots, m_{u}$ takes about $m_{l}(2m_{u}-m_{l})$ 
floating point operations.

\item In order to obtain the $\xi_{IJ}$ coefficient in a subdiagonal row
($I>J$), the number of floating point operations to be performed is $\min\{m_{u},m_{l}
-(I-J)\}=m_{l}-(I-J)$; this is performed in such a way that the total number of floating
point operations related to this row is approximately $m_{l}^{2}$.

\end{itemize}

Finally, obtaining all coefficients $\xi$ would require slightly less operations (due to the
boundary effects) than $2 n m_{u}m_{l}$ floating point operations. Once they are known (or partly
known during the procedure to get them), we can obtain the solution vector $x$ using
the simple recursive relationships presented in this section at a cost of
$4n(m_{u}+m_{l})$ floating point operations.

\section{Banded plus sparse systems}
\label{sec_sparse}

A slight modification of the calculations presented in the previous section is 
required to tackle systems where not all the non-zero entries are within the 
band. The resulting modified procedure is described in this section.

If we have
\begin{equation}
A^\prime := A + \sum_{T=1}^{T_{max}}
  A^\prime_{R_{T}S_{T}}\Delta^{R_{T}S_{T}} \ ,
\end{equation}
with $A$ banded (see eqs.~(\ref{Aup}) and (\ref{Adown})) and the matrix
$\Delta^{R_{T}S_{T}}$ consisting of entries $(\Delta^{R_{T}S_{T}})_{IJ}=
\delta_{I,R_{T}}\delta_{J,S_{T}}$, $\delta_{IJ}$ being the Kroenecker delta,
we shall say that $A^\prime$ is a \emph{banded plus sparse} matrix, and
\begin{equation}
\label{sparse_system}
A^\prime x=b
\end{equation}
a \emph{banded plus sparse} system.
{We call an \emph{extra-band entry} any nonzero entry which does not
lie in the band (this is, $A'_{IJ}$ is an extra-band entry if it is not zero and
$I<J$, $J>I+m_u$ or  $I>J$, $J>J+m_l$). } 

In the pure banded system (section \ref{sec_main}) only $\xi_{K,K+J}$ and $
\xi_{K+I,K}$ with $K=1,\ldots, n$; $J=1, \ldots, m_{u}$; $I=1, \ldots, m_{l}$
had to be calculated. In this case, we also need to obtain
\begin{eqnarray}
 & \xi_{IS_{T}} \quad & \textrm{if $R_{T}<S_{T}$, for $I=R_{T}, R_{T}+1, \ldots, 
  S_{T}-m_{u}-1$} \nonumber \ , \\
 & \xi_{R_{T}J} \quad & \textrm{if $R_{T}>S_{T}$, for $J=S_{T}, S_{T}+1, \ldots, 
  R_{T}-m_{l}-1$} \nonumber \ ,
\end{eqnarray}
with $T=1,\ldots, T_{max}$.

{As seen in the previous section, in order to erase (i.e., turn to 0) 
entry $G_{IJ}$, with $I<J$,
of a generic matrix $G$, we can multiply it by a matrix $P_I$  (see (\ref{defP}, \ref{GP_a}, \ref{GP_b}, \ref{GP_c})).
This action adds the column $I$ (times given numbers) of matrix $G$ to other columns of $G$. This 
erases $G_{IJ}$, but (in general)
adds nonzero numbers to the entries below it ($KJ$ entries with $K>I$).
Therefore, if these entries were zero before performing
the product $GP_I$, they will in general be nonzero after it.
This implies that they will also have to be erased. Hence, erasing the extra-band entry $IJ$ of $A'$ 
will not suffice; the entries $I+1,J$, $I+2,J$, $\ldots$, $J-m_u+1,J$ 
will also have to be erased.
If the extra-band entry to erase $A'_{IJ}$ is below the diagonal ($I>J$), then the $Q_{J}$ matrices
(\ref{defQ}, \ref{QG_a}, \ref{QG_b}) can be used to this end, since they add rows when multiplied by
a generic matrix. Erasing $A'_{IJ}$ will probably make that entries $A'_{IK}$ with $K=J+1,\ldots,I-m_l-1$
become nonzero, and these entries will have to be also erased. 

In order to erase the extra-band entries, the expressions presented in the previous section can be used.
All extra-band entries can lie in an extended band wider than the original band. But, for
the sake of efficiency, the entries in the extended band which are zero during the erasing procedure
must not enter the sums for the coefficients $\xi, \chi$.
}

We define
\numparts
\begin{eqnarray}
\nu_{R_{T}I}  & := \max\{R_{T}, I-m_{l}\} \ , \\
\rho_{S_{T}J} & := \max\{S_{T}, J+m_{u}\} \ .
\end{eqnarray}
\endnumparts
If $R_{T}<S_{T}$ (superdiagonal extra-band entry), in addition to coefficients appearing in 
(\ref{xifinal_a}, \ref{xifinal_b}, \ref{xifinal_c}) we have to calculate
\numparts
\begin{eqnarray}
\xi_{R_{T}S_{T}} & = -\xi_{R_{T}R_{T}} A^\prime_{R_{T}S_{T}}   \ , & \label{spbd1} \\
\xi_{IJ} & = \xi_{II} \left(\sum_{M=\nu}^{I-1}\chi_{IM}\xi_{MJ}\right)
 & \textrm{for $R_{T}<I<S_{T}-m_{u}$} \ ; \label{spbd2}
\end{eqnarray}
\endnumparts
and if $R_{T}>S_{T}$ (subdiagonal extra-band entry), in addition to coefficients appearing in 
(\ref{xifinal_a}, \ref{xifinal_b}, \ref{xifinal_c}) we have to calculate
\numparts
\begin{eqnarray}
\xi_{R_{T}S_{T}} & = -A^\prime_{R_{T}S_{T}}  \ , &  \label{spbd3}\\
\chi_{IJ} & = \sum_{M=\rho}^{J-1}\xi_{IM}\chi_{MJ} 
 & \qquad \textrm{for $S_{T}<J<I-m_{l}$} \ .  \label{spbd4}
\end{eqnarray}
\endnumparts
{The coefficients appearing in (\ref{spbd1}, \ref{spbd2}, \ref{spbd3}, \ref{spbd4}) arise from merely 
applying equations (\ref{xifinal_a}, \ref{xifinal_b}, \ref{xifinal_c}) and avoiding to include in them
the coefficients $\xi$, $\chi$ which are zero due to the structure of $A'$. }

Equations (\ref{spbd1}, \ref{spbd2}, \ref{spbd3}, \ref{spbd4})  have to be modified for $I < J$ if there exist
$A^\prime_{R_{x}S_{T}}$ with $R_{x}<R_{T}$. This is because erasing the
upper non-zero entries by adding columns creates new non-zero entries below
them, and the new relations must take this into account. Analogous corrections
must be done for $I>J$ if there exist $A^\prime_{R_{T}S_{x}}$ with $S_{x} >
S_{T}$. 
{The general rule to proceed in sparse plus banded systems 
is to apply equations (\ref{xifinal_a}, \ref{xifinal_b}, \ref{xifinal_c}) using the maximum $m_u'$,
$m_l'$ so that all the nonzero entries of $A'$ lie within the enhanched band (given by $m_u'$, $m_l'$),
and avoid that the coefficients ($\xi$, $\chi$) which are zero take part in the sums. 
The coefficients $\xi_{KL}$ which are zero are those given by the following rules:
\begin{itemize}
\item If $K<L$, $\xi_{KL}=0$ if $A'_{ML}=0$ for $M=1,\ldots,K$
\item If $K>L$, $\xi_{KL}=0$ if $A'_{KM}=0$ for $M=1,\ldots,L$
\end{itemize}
}
The computational cost of solving banded plus sparse systems scales with $n$, as long
as the number of columns above the band and rows below it containing non-zero
entries $A^\prime_{R_{T}S_{T}}$ is small ($\ll n$). {
The example code for an algorithm for sparse plus 
banded systems can be found in the supplementary material; the performance of this algorithm is 
presented in sec. \ref{sec_application}.
}


\section{Algorithmic implementation}
\label{algorithms}

Based on the expressions (\ref{xifinal_a}, \ref{xifinal_b}, \ref{xifinal_c},
\ref{cfinal}, \ref{xfinal}) derived in the paper, we have coded several different algorithms
that efficiently solve the linear system in (\ref{basic_system}). The
difference between the method in this paper and the most commonly used
implementation of Gaussian elimination techniques, such as the ones included
in LAPACK \cite{And1999BOOK}, Numerical Recipes in C \cite{Pre2007BOOK}, or
those discussed in ref.~\cite{WatXXXXBOOK} is that these methods perform an
$LU$ factorization of the matrix $A$, and the coefficients $\xi$ for the
Gaussian elimination are obtained in several steps, whereas the method
introduced here does not perform such an $LU$ factorization, and it obtains the
coefficients $\xi$ in a single step.

In order to obtain the solution of (\ref{basic_system}) we need to get the
coefficients $\xi$ for Gaussian elimination as explained in section
\ref{basis}. That is, one diagonal coefficient for each row/column, plus $m_{u}$
coefficients in each row and $m_{l}$ coefficients in each column (except for
the last ones, where less coefficients have to be calculated). More accuracy
 in the solution is obtained by pivoting, i.e., altering the order
of the rows and columns in the process of Gaussian elimination so that the
\emph{pivot} (the element temporarily in the diagonal and by which we are
going to divide) is never too close to zero. Double pivoting (in rows
\emph{and} columns) usually gives more accurate results than partial pivoting
(in rows \emph{or} columns). However, the former is seldom preferred for
banded systems, since it requires $\mathcal{O}(n^{2})$ operations, while the
latter requires only $\mathcal{O}(n)$. In the implementations described in
this section, we have chosen to perform partial pivoting on rows, as in
refs.~\cite{Pre2007BOOK,And1999BOOK}. In the same spirit, and in order to
save as much memory as possible, we store matrices by diagonals (see \cite{Pre2007BOOK}).

We proceed as follows: For each given $I$, we obtain $\xi_{II}$ (using
(\ref{xifinal_a})), and then $\xi_{JI}$ (using (\ref{xifinal_c})) for $J=I+1,
\ldots, I+m_{l}$. If $|\xi_{JI}|>|\xi_{II}|$, we exchange rows $I$ and $J$ in
the matrix $A$ and in the vector $b$. This is called partial pivoting in rows,
and it usually gives greater numerical stability to the solutions; in our
tests of section \ref{rad} the error was lowered in two orders of magnitude by
partial pivoting. Next, we calculate $\xi_{IJ}$ (using (\ref{xifinal_b})) for
$J=I+1, \ldots, I+m_{u}$. When we have calculated all the relevant
coefficients $\xi$ for a given $I$, we calculate $c_I$ using (\ref{cfinal}).
We repeat these steps for all rows $I$, starting by $I=1$ and moving one row
at a time up to $I=n$. This ordering enables us to solve the system using
eqs.~(\ref{xicoefsfinal}), (\ref{cfinal}), (\ref{xfinal}), because the
superdiagonal $\xi_{IJ}$ (i.e., those with $I<J$) only require the knowledge
of the coefficients with a lower row index $I$, while the subdiagonal
coefficients $\xi_{IJ}$ with $I>J$ only require the knowledge of coefficients
with a lower column index. We have additionally implemented a procedure to
avoid performing dummy summations (i.e., those where the term to add is null),
which eliminates the need for evaluating $\mu_{IJ}$ in every step. According
to the pivotings performed before starting to calculate a given $\xi$, a
different number of terms will appear in the summation to obtain it. This
procedure uses the previous pivoting (i. e., row exchanging) information and
determines how many $\xi$ coefficients have to be obtained in any row or
column, and how many terms the summation to obtain them will consist of (this
procedure is not indicated in the pseudo-code below for the sake of
simplicity). The final step consists of obtaining $x$ from $b$ using
(\ref{xfinal}).

The pseudo-code of the algorithm can be summarized as follows:

\begin{algorithmic}
\STATE
\STATE{// Steps 1 and 2: Calculating the coefficients $\xi$ and the vector 
          $c$}
\FOR{($K=1, \ K \leq n, \ K++$)}
\STATE{  // Calculating the diagonal $\xi$'s:}
\STATE{$\xi_{KK}=1/(A_{KK}+\sum_{M=K-\mu'}^{K-1}\xi_{KM}\xi_{MK} )$}
\STATE{  // Calculating the subdiagonal $\xi$'s:}
\FOR{($I=K+1, \ I \leq I+m_{l}, \ I++$)}
\STATE $\xi_{IK}=-A_{IK}+\sum_{M=K-\mu_{IK}}^{K-1}\xi_{IM}\xi_{MK}  \qquad 
       \textrm{for } I>J$
\ENDFOR
\STATE
\STATE  // Pivoting: 
\IF {$\exists \ |\xi_{JI}|>|\xi_{II}|$ for $J=I+1, \ldots, I+m_{l}$}
	\FOR{($K=1, \ K \leq n, \ K++$)}
		\STATE $A_{IK} \leftrightarrow A_{JK} $
	\ENDFOR
	\STATE $b_{I}\leftrightarrow b_{J}$
\ENDIF
\STATE
\STATE{   // Calculating the superdiagonal $\xi$'s:}
\FOR{($J=K+1, \ J \leq K+m_{u}, \ J++$)}
\STATE $\xi_{KJ}=-\xi_{KK}A_{KJ}+\sum_{M=K-\mu_{KJ}}^{K-1}\xi_{KM}\xi_{MJ}$
\ENDFOR
\STATE
\STATE{   // Calculating $c_{K}=(Qb)_{K}$:}
\STATE{$c_{K}=b_{K}$}
\FOR {($L=K-m_{l}, \ L \leq K - 1, \ L++$)}
\STATE{$c_{K}+=\xi_{K,L}c_{L}$}
\ENDFOR

\ENDFOR 
\STATE

\STATE{// Step 3: Calculating $x_{K}=(Pc)_{K}=(PQb)_{K}$:}
\FOR{($K=n, \ K \geq 1, \ K--$)}
\STATE{$x_{K}=\xi_{KK}c_{K}$}
\FOR {($L=K+1, \ L \leq K+m_{u}, \ L++$)}
\STATE{$x_{K}+=\xi_{K,L}x_{L}$}
\ENDFOR
\ENDFOR
\STATE
\end{algorithmic}

In the actual computer implementation we split the most external loop into
three loops ($I=1, \ldots, 2m$, $I=2m+1, \ldots, n-2m$, and $I=n-2m+1, \ldots,
n$), because the summations to obtain the coefficients $\xi$ lack some terms
in the initial and final rows. We store $A$ by diagonals in a $n \times
(2m_{u}+m_{l}+1)$ matrix in order to save memory and, with the same objective,
we overwrite the original entries $A_{IJ}$ with the calculated $\xi_{IJ}$ for
$I \leq J$, and we store the $\xi_{IJ}$ with $I > J$ in another $n \times
(2m_{l})$ matrix.

One possible modification to the algorithm presented above is to omit the
pivoting. This usually leads to larger errors in the solution, but results in
important computational savings. It can be used in problems where
computational cost is more important than achieving a very high accuracy. In
any case, one must note that the accuracy of the algorithm is typically
acceptable without pivoting, so in many cases no pivoting will be necessary.

In (\ref{xi}) we can see that no subdiagonal coefficients ($\xi_{IJ}$ with
$I>J$) are needed to obtain $x$ from $c$. In (\ref{ci}), we can see that only
$\xi_{IK}$ are necessary in order to obtain $c_{I}$, thus making it
unnecessary to know $\xi_{LK}$ for $L<I$. Therefore, we can get rid of them
once $c_{I}$ is known. Since we calculate $c_{I}$ immediately after
calculating all $\xi_{IK}$, we can overwrite $\xi_{I+1,K}$ on the memory
position of $\xi_{IK}$. If we do so, about one third of the memory is saved,
since less coefficients must be stored, however, according to some preliminary
tests, this option is also 20\% slower than the simpler one in which all
coefficients are stored independently.

It is also worth remarking at this point that the present state of the
algorithm is not yet completely optimized at the low level and, therefore,
it cannot be directly compared to the thoroughly optimized routines included
in commonly used scientific libraries such as LAPACK \cite{And1999BOOK}. This
further optimization will be pursued in future works.

\section{Parallelization}
\label{sec_parallelization}

There exist many works in the literature aiming at parallelizing the calculations
needed to solve a banded system 
\cite{Don1987PC,Pol2006PC,Cha2003SIAM,Bin1999SIAM,Mei1985PC,Zha1994PC,Law1984ACM,Joh1985ACM,Che1978ACM,Eva1976CJ,Gar2000AMC,Arb1994TR,Gol2001NLAA,Don1984PC}. 
The decision about which one to choose, and, in particular, which one to apply to the algorithms
presented in this work depends, of course, on the architecture of the machine
in which the calculations are going to be performed. The choice is
additionally complicated by the fact that, normally, only the number of
floating point operations required by each scheme is reported in
the articles. However, the number of floating point operations is known to be a poor measure of
the real wall-clock performance of computer algorithms and, especially,
parallel ones, for a number of reasons:

\begin{itemize}

\item Not all the floating point operations require the same time. For example, in currently
common architectures, a quotient takes 4 times as many cycles as an addition
or a product.

\item A floating point operation usually requires access to several positions of memory. Each
access is much slower than the floating point operation itself \cite{HenXXXXBOOK}. Moreover, the
number of memory accesses does not need to be proportional to the number of
floating point operations.

\item Transferring information among nodes in a cluster is commonly much
slower than accessing a memory position or performing a floating point
operation \cite{HenXXXXBOOK}.

\end{itemize}

Despite these unavoidable complexities and the fact that rigorous tests should
be made in any particular architecture, two parallelizing schemes seem well
suited for the method presented in this work: the one in ref.~\cite{Mei1985PC}
for shared-memory machines and the one in ref.~\cite{Pol2006PC} for
distributed-memory machines. The former is faster if the communication time
among nodes tends to zero, whereas the latter tackles the communication time
problem by significantly reducing the number of messages that need to be 
passed.

\section{Differences with Gaussian elimination}
\label{rad}

In order to asess the performance of the method derived in the previous sections, 
we will present results of numerical tests of real systems. In sec. \ref{secalltests},
we compare the absolute accuracy and
numerical efficiency of our New Algorithm with those of the banded solver
described in the well-known book \emph{Numerical Recipes in C}
\cite{Pre2007BOOK}. 
However, before doing that, we can make some general remarks about the validity of the 
new method from the numerical point of view. 

At this point, it is worth remarking that the present state of our New
Algorithm for banded systems is not yet completely optimized at the low level. Therefore,
it cannot be directly compared to the thoroughly optimized routines included
in commonly used scientific libraries such as LAPACK \cite{And1999BOOK}. This
further optimization will be pursued in future works.
At the current state, it is natural to compare our algorithm to an explicit, 
high-level, 
not optimized routine, such as the ones in \emph{Numerical Recipes in C}
\cite{Pre2007BOOK}, and the results here should be interpreted as a hint
of the final performance when all levels of optimization are tackled.

Our New Algorithm (NA) is based on equations 
(\ref{xifinal_a2}, \ref{xifinal_b2}, \ref{xifinal_c2}), (\ref{cfinal}) and 
(\ref{xfinal}). The source code of its different versions can be found in the supplementary material.
The solver of \cite{Pre2007BOOK} (NRC)
belongs to a popular family of algorithms (see, for example, \cite{WatXXXXBOOK})
which work by calculating the $\xi$
coefficients involved in the Gaussian elimination procedure in different iterations.
Both for NA and for NRC, the $\xi$ coefficients required for the resolution result from the summation of several terms. 
For a given set of $\xi$'s, Gaussian elimination-based methods
first obtain the first term of the summation of every $\xi$ then the corresponding second
terms of the summations, and so on. In contrast, our method first obtains the final value of a
given $\xi$ by calculating all the terms in the corresponding summation;
then once a given $\xi_{IJ}$ is known, it computes $\xi_{I,J+1}$, and so on.

{
Both the NRC Gaussian elimination method for banded systems and our New Algorithm perform the same
number of operations (i.e., the same number of additions, the same number of products, etc.). 
However, their efficiencies are different, as is shown in sec. \ref{numtests}.
We believe this is due to the way the computers which run the algorithms access the memory positions
which store the variables
involved in the problem. The time that modern computers take to perform a 
floating point operation with two variables can be much smaller than the time required to access 
the memory positions of these two variables. In a modern computer, an addition or product of
real numbers can take of the order of $10^{-9}$-$10^{-8}$ s. If the variables involved are 
stored in the cache memory, to access them can take also $10^{-9}$-$10^{-8}$ s. If
they are stored in the main memory, the access can take the order of $10^{-7}$ s 
\cite{libro_HPC,HenXXXXBOOK}. The speed to access 
one position of memory is given not only by the level of memory (cache, main memory, disk, etc.)
where it lies, but also by the proximity of the position of memory which was immediately read 
previously. Therefore, it is expected that two different algorithms will require
different execution times if they access the computer memory in different ways, even if they perform
the same operations with the same variables. 

In the NRC Gaussian elimination procedure,
a given number of floating point variables
is added to each entry $A_{IJ}$. The same number of floating point variables has to
be added to $A_{IJ}$ to calculate $\xi_{IJ}$ in the New Algorithm. However, the \emph{order} it is
done is different in both cases. In Gaussian elimination, one row of $A$ times a given number is 
added to another row of $A$, and this is repeated many times. For example, for erasing the subdiagonal
entries of the first column of $A$, the first row (times the appropriate numbers) is added to rows
2 to $m_l+1$. Then, to erase the (new) second column, the (new) second row is added to rows 3 to $m_l+2$, and so on.
Let us consider $A_{4,3}$, without a loss of generality.
A given number is added to $A_{4,3}$ when the first row is added to
its lower rows; after some steps, another number is added to $A_{4,3}$, when the second row is added to
its lower rows. Again after some steps, another number is added to $A_{4,3}$ (when the third row is added to
its lower rows). In this procedure, the memory positions that are accessed move away from the position of $A_{4,3}$,
and then they come back to it, which can be suboptimal. However, in our New Algorithm,
numbers are added to a memory position (say $A_{4,3}$) only once (see \ref{xifinal_c}),
making the sweeping of memory positions more efficient. Some simple tests seem to support this 
hypothesis. We define the following loops:
\begin{itemize}
\item Loop 1: (Analogous to loop of NRC-Gaussian elimination)
\begin{algorithmic}
\FOR{($K=0, \ K < 1000000, \ K++$)}
\FOR{($I=0, \ I < m_l, \ I++$)}
\FOR{($J=0, \ J < m_u-1, \ J++$)}
\STATE{  $A[m_l][J]+=A[I][J]$}
\ENDFOR
\ENDFOR
\ENDFOR
\end{algorithmic}
\item Loop 2: (Analogous to the loop of the New Algorithm)
\begin{algorithmic}
\FOR{($K=0, \ K < 1000000, \ K++$)}
\FOR{($I=0, \ I < m_u-1, \ I++$)}
\STATE{  $A[m_l][I]+=A[0][I]+\ldots+A[m_l-2][I]$}
\ENDFOR
\ENDFOR
\end{algorithmic}
\end{itemize}
The way in which Loop 1 sweeps the memory positions is analogous to that of the Gaussian elimination method (NRC), 
because it adds the different numbers to a given memory position ($A[m_l][I]$) in different iterations within the 
intermediate loop.
The way Loop 2 sweeps the memory positions is analogous to that of the New Algorithm (NA), 
because it adds the different numbers to a given memory position ($A[m_l][I]$) just at a stretch.
If we compare the times required by their executions (using $m_u=10$), we find the results of table \ref{tablevs}.

\begin{table}[h]
\caption{{\label{tablevs} Comparison of the execution times of simple tests. The last column corresponds
to data taken from sec. \ref{secalltests}.}
}
\begin{indented}
\item[]\begin{tabular}{@{}cccc|c}
\br
$m_l$ & $t_{\mathrm{Loop 1}}$ & $t_{\mathrm{Loop 2}}$ & $t_{\mathrm{Loop 1}}/t_{\mathrm{Loop 2}}$ & $t_{NRC}/t_{NA}$   \\ 
\mr
3	& 0.379	& 0.169	& 2.243	& 	1.145 \\ 
10	& 1.255	& 0.509	& 2.466	& 	1.780  \\ 
30	& 3.788	& 1.473	& 2.563	& 	2.360  \\ 
\br
\end{tabular}
\end{indented}
\end{table}
\noindent{This simple test gives us a clue on how the different ways to sweep memory positions can result in rather 
different execution times. The comparison between 
$t_{\mathrm{Loop 1}}/t_{\mathrm{Loop 2}}$ and $t_{NRC}/t_{NA}$
associated with the actual NRC and NA algorithms (without pivoting, with $N=10^6$ and $m_u=m_l$, see
sec. \ref{secalltests})
is merely qualitative. This is because the way the memory access takes place in Loop 1 is not 
exactly the same as the way the memory access takes place in NRC, nor is it the same for Loop 2 and NA 
(and also the $m_u$'s are different).
The reason why the relative performance of NA vs. NRC decreases with $m$ for $m>35$ approximately can be due to the fact
that in our implementation, NA uses matrices which are larger than those of NRC.}
}


\section{Numerical tests}\label{secalltests}

In this section we quantitatively compare our New Algorithm with
the banded solver based on Gaussian elimination of \cite{Pre2007BOOK}.
We do so by comparing the accuracy and efficiency of both algorithms for solving given systems.
For the sake of generality, in the first part of this section (\ref{numtests})
we use random banded systems as inputs for our tests. In the second part (\ref{sec_application}), 
we use them (plus a modified version of NA) to solve a physical problem, the calculation of the 
Lagrange multipliers in proteins.

\subsection{Performance for generic random banded systems}\label{numtests}
In the test systems for our comparisons we imposed $m_{u}=m_{l}=m$ for the sake of
simplicity. We took $n=10^{3}, 10^{4}$ and $10^{5}$ and $m=3,10,30,100$ and
$300$ for all of them in our tests. In addition to this, we took $n=10^{6}$ with
$m=3, 10$ and $30$. For each given pair of values of $n$ and $m$, we generated
a set of $1000$ random $n \times n$ banded matrices whose
entries are null, except the diagonal ones and their first $m$ neighbours on
the right and on the left. The value of these entries is a random number
between $500$ and $-500$ with 6 figures. The components of the independent
term (the vector $b$ in (\ref{basis})) are random numbers between 0 and 1000,
also with 6 figures. We tested both algorithms (from \cite{Pre2007BOOK}
and our NA) with and without pivoting. We used PowerPC 970FX 2.2 GHz machines,
and no specific optimization flags were given to the compiler apart from the basic one 
(g++ -o solver solver.cpp). Every
point in our performance plots corresponds to the mean of 1000 tests and, in
each point, we used the same random system as the input for both algorithms.

We measured the efficiency of a given algorithm by using an average of its execution times for given
banded systems.
In the measurement of such execution times, we considered only the
computation time;
{ i.e., the measured execution times correspond only to the solution of the banded systems,
and not to other parts of 
the code such as the generation of random matrices and vectors, the
initialization of variables or the checking and the storage of the results.
The measurements of the execution times start immediately 
after the initialization, and the clock is stopped immediately after the unknowns $x$ are calculated. 
The concrete information on how the measurement of 
times was implemented can be can be found in the source code of the programs used for our tests, which are 
included in the supplementary material. As is shown there, 
standard C libraries were used to measure the times.}

The accuracy of the algorithms was determined by measuring the error of the solutions they provide ($x$).
We quantified the error with the following formula, which
corresponds to the normalized deviation of $Ax$ from $b$:
\begin{equation}
\label{Err}
\mathrm{Error}:=\frac{\sum_{I=1}^{n}|\sum_{J=1}^{n}A_{IJ}x_{J}-b_{I}|}
       {\sum_{I=1}^{n}|x_{I}|} \ .
\end{equation}

\begin{center}   
\begin{figure}[!ht]
\includegraphics[width=13.5cm]{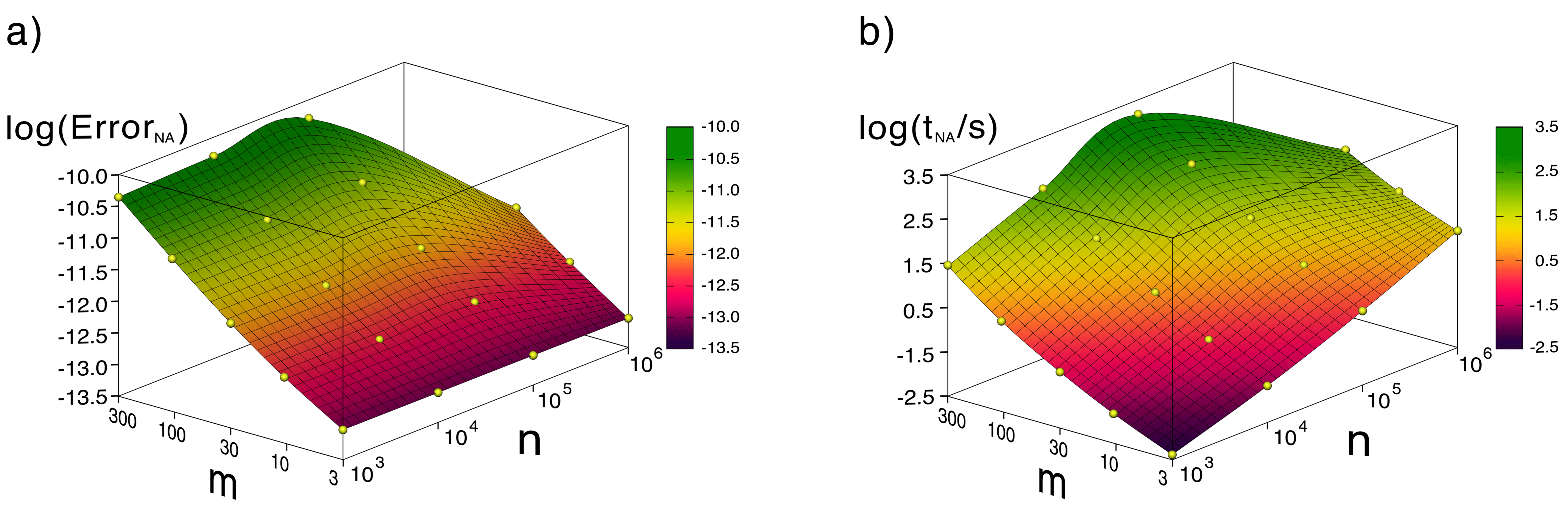}
\caption{\small{Properties of the New Algorithm introduced in this work, with
 pivoting, as a function of the size of the matrix $n$ and the width of the 
 band $m$ in random banded test systems. \textbf{a)} Its accuracy, as 
 measured by the error defined in equation~(\ref{Err}). \textbf{b)} Its numerical 
 efficiency, measured by the execution time.}}
\label{pivabs}
\end{figure}
\end{center}

\begin{figure}[!ht]
\begin{center}     
\includegraphics[width=13.5cm]{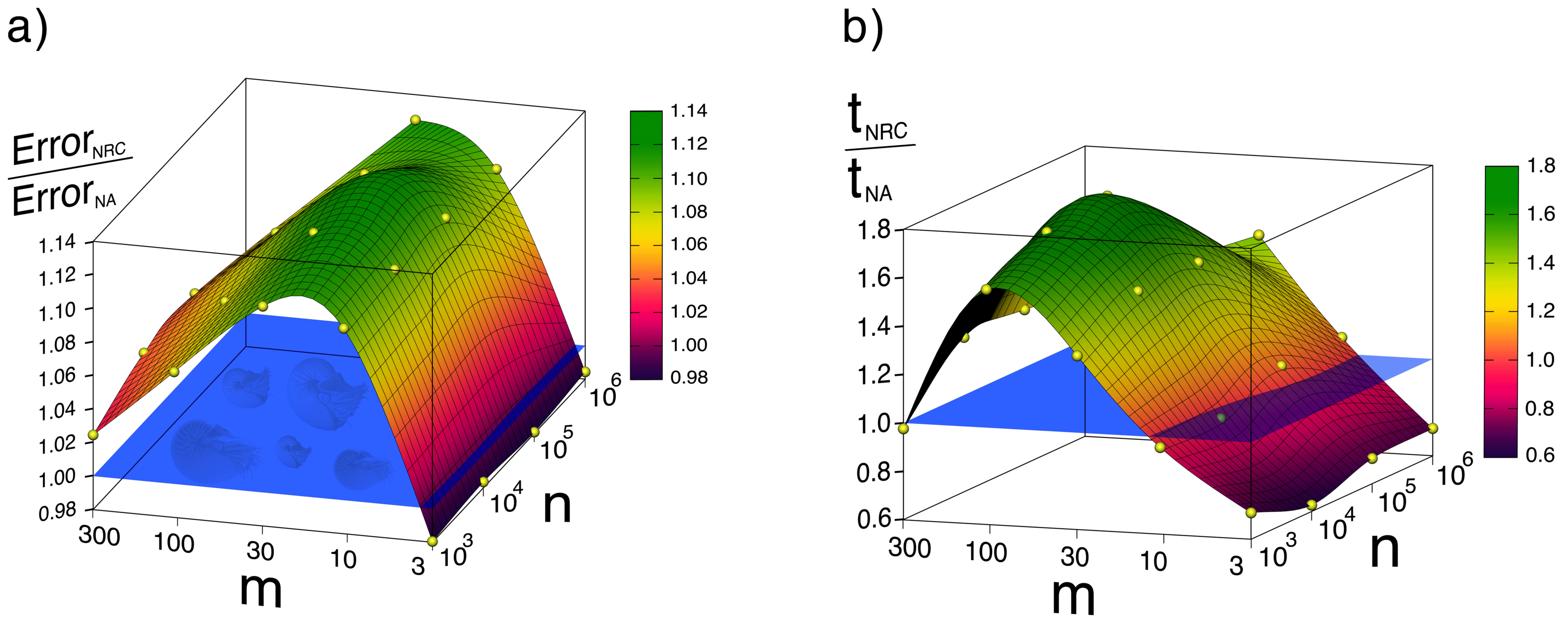}
\caption{\small{Comparison between the properties of the New Algorithm 
 introduced in this work (NA) and the one in ref.~\cite{Pre2007BOOK} (NRC),  
 both with pivoting, as a function of the size of the matrix $n$ and the width 
 of the band $m$ in random banded test systems. \textbf{a)} Relative accuracy, 
 as measured by the ratio of the errors defined in equation~(\ref{Err}). 
 \textbf{b)} Relative numerical efficiency, measured by the ratio of the 
 execution times.}}
\label{pivcomp}
\end{center}
\end{figure}

\begin{figure}[!t]
\begin{center}     
\includegraphics[width=9.5cm]{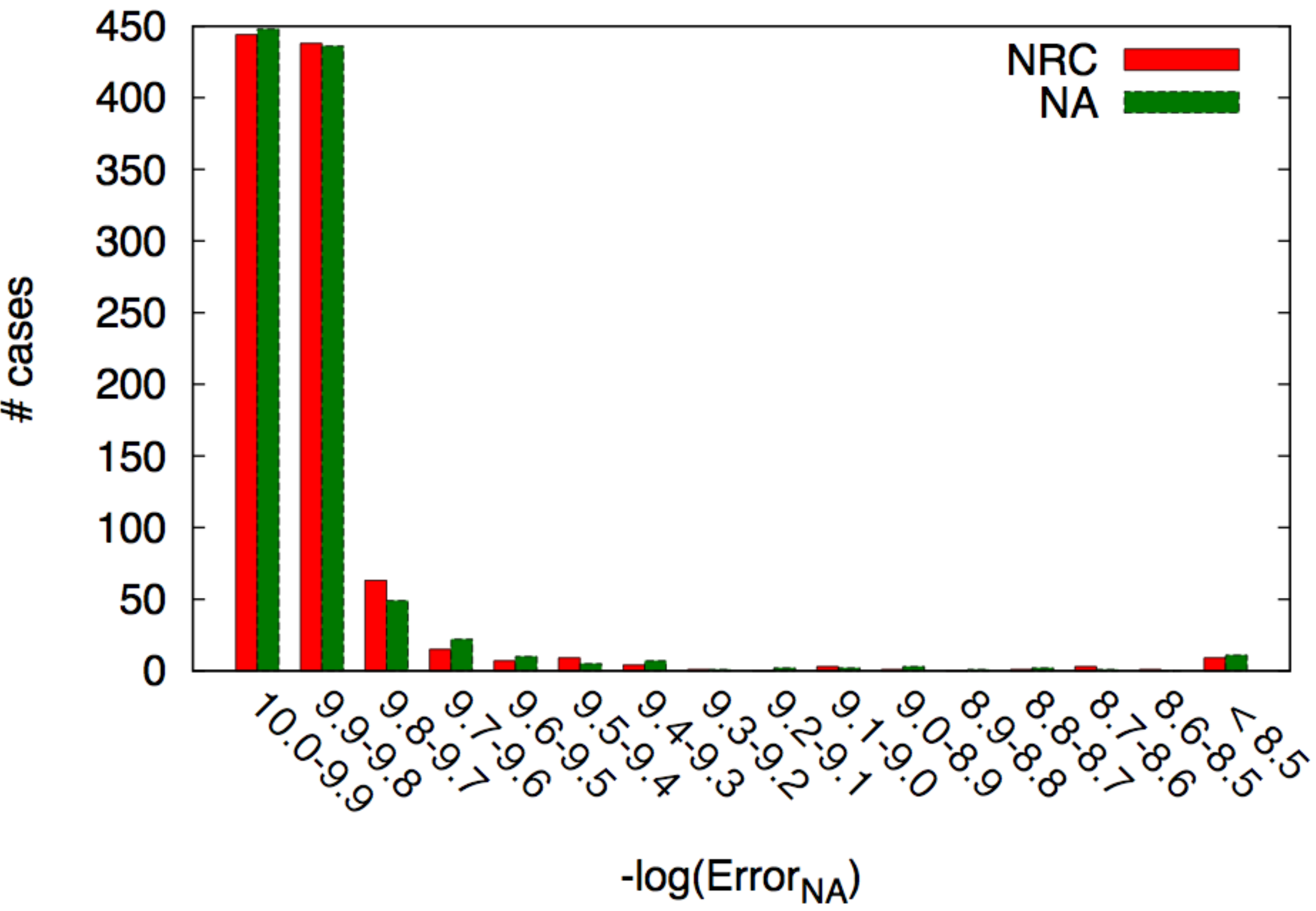}
\caption{\small{Histogram of the errors made by the algorithms NA and NRC for 
 solving random banded systems without pivoting. The data corresponds to 
 1000 random inputs with $n=10^{5}$ and $m=10$.}}
\label{histogr_err_NE5_m10}
\end{center}
\end{figure}

\begin{figure}[h]
\begin{center}     
\includegraphics[width=13.5cm]{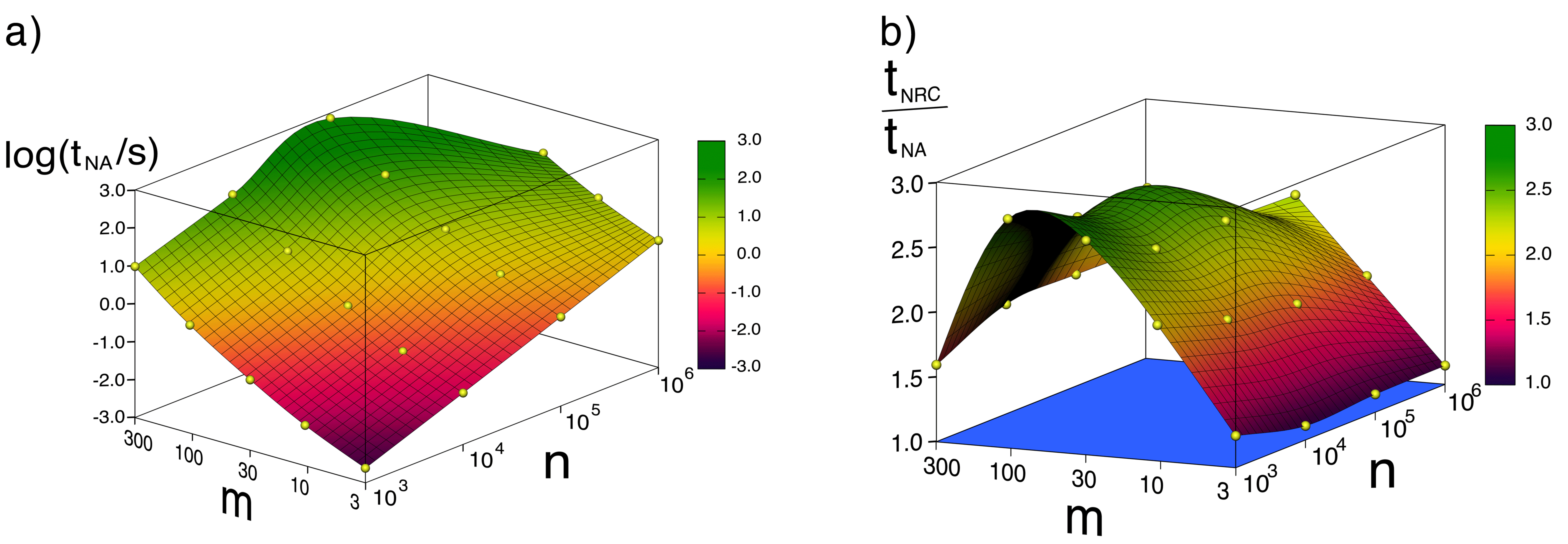}
\caption{\small{\textbf{a)} Numerical complexity of the New Algorithm
 introduced in this work, without pivoting, when solving random banded test 
 systems. Execution time is shown as a function of the size of the matrix $n$ and the
 width of the band. \textbf{b)} Comparison between the numerical complexity
 of NA and NRC.}}
\label{nopiv}
\end{center}
\end{figure}

In the first diagrams of this section (figs. \ref{pivabs}, \ref{pivcomp}, \ref{histogr_err_NE5_m10} and \ref{nopiv})
we present the absolute and relative accuracy and
efficiency of the NA and NRC algorithms
for the cases with and without pivoting. In these figures, the yellow
spheres represent the calculated points, which correspond to the
average of 1000 tests with different input random matrices and vectors. For
the sake of visual confort, interpolating surfaces have been produced with
cubic splines and the $x$ and $y$ axes (labeled $n$ and $m$) are in logarithmic scale. In figures
\ref{pivcomp}, \ref{nopiv} we compare quantities between the two algorithms;
a blue plane at $z=1$ is included. Above this plane, NA is more competitive than NRC; below this plane, 
the converse is true.

In figure \ref{pivabs}a, we can see that our algorithm with pivoting has 
very good accuracy, with the error satisfying $\mathrm{log}(\mathrm{Error}) \propto
\mathrm{log}(m)$. The error is proportional to a power of $m$
with a small exponent ($\simeq 1.4$). In the same figure we notice that 
this error is approximately independent of $n$.
The execution time in
the tested region (see figure~\ref{pivabs}b) is proportional to $n$ and also
approximately proportional to $m^{1.7}$, not to $m^{2}$ as one would expect
from the number of floating point operations ($\propto n m^{2}$). This suggests that memory access
is an important time-consuming factor, in addition to floating point
operations.

In figure \ref{pivcomp}a, we can see that, if pivoting is performed, our New
Algorithm is always more accurate than NRC, except for a narrow range of $m$
between 1 and 4. The typical increase in accuracy is around a 5\%, reaches
almost 15\% for some values of $n$ and $m$. In figure \ref{pivcomp}b, we can
see that, if pivoting is performed, the New Algorithm is also faster than NRC
for most of the studied values of $n$ and $m$, with typical speedups of
around 40\% and the largest ones of almost 80\%.

If pivoting is not performed, the accuracy decreases typically by two or three
orders of magnitude (but errors still remain very low, usually around
$10^{-10}$).In the non-pivoting case, we also see that a few of the
calculations (around 1 in 500) present errors significantly larger than the
average. This probably suggests that the random procedure has produced a matrix
that is close to singular with respect to the hypotheses introduced in
sec.~\ref{sec_main}. In fig. \ref{histogr_err_NE5_m10}, we show a typical example of the distribution of errors
for the non-pivoting banded solvers NA and NRC in
figure~\ref{histogr_err_NE5_m10}. The data corresponds to the errors of 1000
random input matrices with $n=10^{5}$ and $m=10$. In such a case, the average
of the error is less representative. In this example test, the highest error
in NRC is $1.57 \cdot 10^{-9}$, and in NA is $2.53\cdot 10^{-9}$, although
these numbers are probably anecdotal. One must also note that $99\%$ of the
errors are $\mathcal{O}(10^{-10})$ or smaller. A comparison of the red and
green bars in the histogram suggests that there are no big differences in the
errors of both algorithms (NA and NRC) without pivoting.

Despite these problems in dealing with almost singular matrices, algorithms without pivoting
have an important advantage regarding computational cost, and they can be useful
for problems in which the matrices are a priori known to be well behaved.
These computational savings are noticed if we compare figs.~\ref{pivabs}b, 
and~\ref{nopiv}a. In figure~\ref{nopiv}b, we can additionally see that
the New Algorithm introduced in this work is always faster than NRC for the
explored values of $n$ and $m$ if no pivoting is performed; the increase in
efficiency reaching almost to a factor of 3 for some values of $n$ and $m$, and
being typically around a factor of 2.

\subsection{Analytical calculation of Lagrange multipliers in a protein}\label{sec_application}
{
In Molecular Dynamics simulations, it is a common practice to \emph{constrain} 
some of the internal degrees of freedom of the involved systems. This enables an increase in the 
simulation time step, makes the simulation more efficient, and is
expected not to severely distort 
the value of the observable quantities calculated in the simulation \cite{Leimkuhlerbook,Hes1997JCC}.
The bond lengths of a 
molecule can be constrained by including 
algebraic restrictions such as the following one:
\begin{equation}\label{sigma_generica}
|\vec{x}_{\alpha}-\vec{x}_{\beta}|^{2}-(a_{\alpha,\beta})^{2} = 0 \ 
\end{equation}
in the system of classical equations of motion of the atoms.
In this expression, the positions of atoms in a molecule formed by $N_a$ atoms 
are given by $\vec{x}_{\alpha}$, $\vec{x}_{\beta}$,
with $\alpha, \beta = 1, \ldots, N_a$. The parameter $a_{\alpha,\beta}$ is the length of the bond
which links atoms $\alpha$ and $\beta$.

The imposition of holonomic constraints such as (\ref{sigma_generica}) under the assumption of the 
D'Alembert principle
makes the so-called \emph{constraint forces} appear. These
forces are proportional to their associated \emph{Lagrange multipliers}, which have to be 
calculated in order to evaluate the dynamics of the system.
Proteins, nucleic acids and other biological molecules have an essentially linear topology,
which makes it 
possible to calculate the Lagrange multipliers associated to their constrained internal degrees of freedom
by solving banded systems. More explanations on
how to impose constraints on molecules and on how to calculate the Lagrange multipliers 
in biomolecules can be found in \cite{GR2011JCC}. 

In this section, we compare the efficiencies and accuracies of three methods to solve the banded systems
associated with the calculation of Lagrange multipliers of a family
of relevant biological molecules (polyalanines). 
The three methods we compare are:
\begin{itemize}
\item The Gaussian elimination algorithm for banded systems presented in \cite{Pre2007BOOK} (NRC)
\item The New Algorithm (NA) presented here, based on equations (\ref{xifinal_a2}, \ref{xifinal_b2}, \ref{xifinal_c2})
\item A modified version of the New Algorithm presented here, which 
uses the methods discussed in sec. \ref{sec_sparse} and
takes advantage in the symmetry of
the system (i.e., it uses equation (\ref{xifinal_csim}) instead of (\ref{xifinal_c2}))
\end{itemize}
All three methods are implemented without pivoting. The accuracies and efficiencies of the 
first two ones were compared in sec. \ref{numtests} for banded matrices with random entries.

In our tests, we calculated the Lagrange multipliers of $\alpha$-helix shaped polyalanine chains (as the one
displayed in fig. \ref{fig:polyalanine}) with different numbers of residues (R).
See \cite{GR2011JCC} for further information on the way the systems of equations to solve were generated.
In our tests, we measured the error as calculated with (\ref{Err}), as well as the execution time of the algorithms.
We ran them in a MacBook6,1 with a 2.26 GHz Intel Core 2 Duo processor.

\begin{figure}[!h]
\begin{center}     
\includegraphics[width=15.6cm]{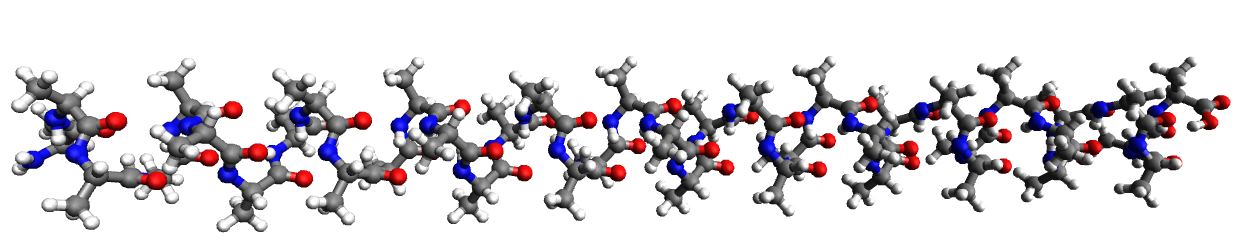}
\caption{{\small{Polyalanine chain of 40 residues in a $\alpha$-helix shape. 
White spheres indicate H atoms, dark spheres indicate C atoms, blue spheres indicate N atoms and red spheres indicate O atoms.
The covalent bonds appear as rods connecting them. Diagram made with Avogadro \cite{Avogadro}.}}}
\label{fig:polyalanine}
\end{center}
\end{figure}

The results are displayed in figures \ref{fig:comppolt} and \ref{fig:comppolerr}. 
\begin{figure}[!h]
\begin{center}     
\includegraphics[width=11.1cm]{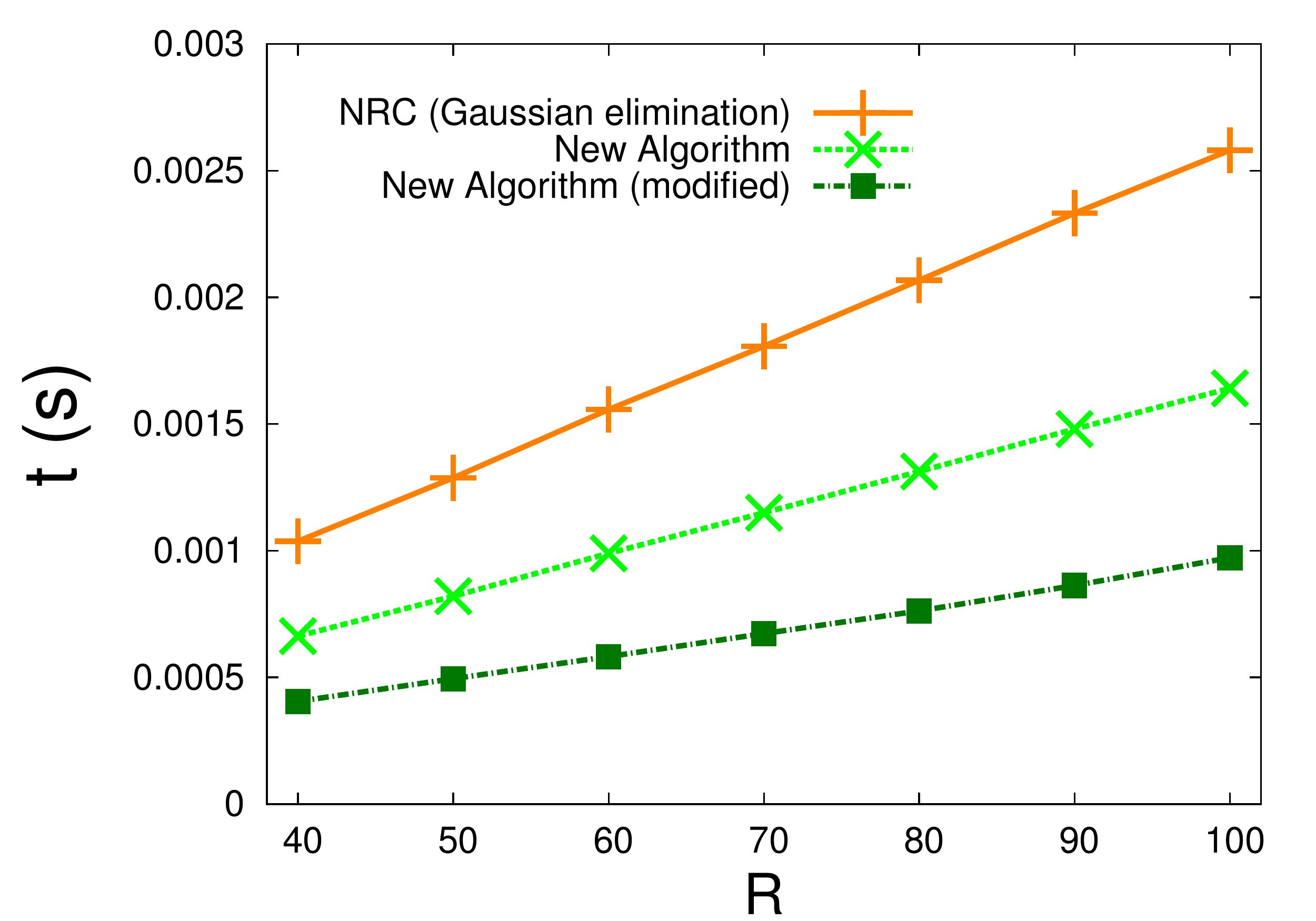}
\caption{{\small{Comparison of the execution times (t) of different algorithms to solve banded systems
in the calculation of the Lagrange multipliers in a polyalanine chain of R residues. Vertical crosses: 
NRC (Gaussian elimination); Diagonal crosses: NA; Squares: modified New
Algorithm.}}
}
\label{fig:comppolt}
\end{center}
\end{figure}
For all the polypeptide lengths represented in fig. \ref{fig:comppolt}, the execution time of the Gaussian elimination
algorithm (NRC) is about 1.57 times the execution time of the New Algorithm ($1.57 \pm 0.01 $). The modified
New Algorithm (squares in figures \ref{fig:comppolt} and \ref{fig:comppolerr}) is about 2.70 
times faster than the NRC
algorithm. These results were the expected results for the used values of $n$, $m_u$ and $m_l$ 
($m_u=m_l=m=6$, $n=10\mathrm{R}+2$), according to the tendencies observed in the previous
section.
Higher values of $m$ are expected to result in better relative efficiency of the New Algorithm (see 
sec. \ref{numtests}). A situation that we can meet,
for example, if not only bond lengths, but also bond 
angles, are constrained, and if the branches of the molecule are longer (for example,
the side chains of the arginine residue are longer than the side chains of the alanine
residue).

The errors made by the three tested algorithms are displayed in fig. \ref{fig:comppolerr}.
\begin{figure}[!h]
\begin{center}     
\includegraphics[width=11.1cm]{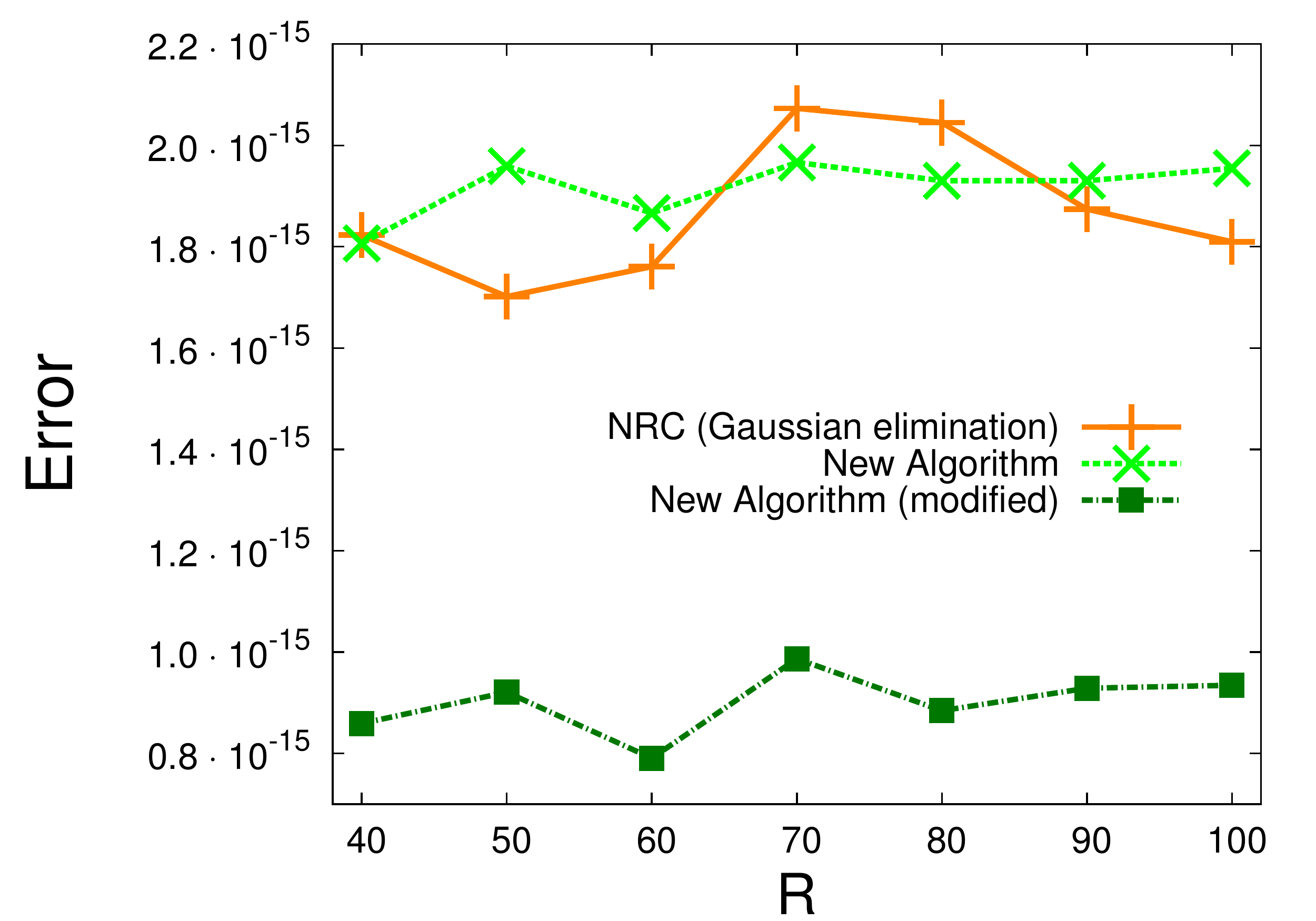}
\caption{{\small{Comparison of the errors made by different algorithms in the solution of banded systems
for the calculation of the Lagrange multipliers in a polyalanine chain of R residues.  
Vertical crosses: 
NRC (Gaussian elimination); diagonal crosses: NA; squares: modified New
Algorithm.}}}
\label{fig:comppolerr}
\end{center}
\end{figure}
As expected for the case without pivoting (see sec. \ref{numtests}), the errors of the NRC and NA algorithms
are similar, and both are very small (similar to the errors arisen from the finite
machine precision). The error of the modified 
version of the New Algorithm is typically less that half of the error of the other two methods. This
can be due to the fact that the modified version uses equation (\ref{xifinal_csim})
instead of (\ref{xifinal_c2}). Therefore, fewer numbers (about half of them)
are present in the calculation, and hence
fewer potential sources of error are present.
}

\vspace{1cm}

We conclude that, for the systems tested in this section,
the new algorithm introduced in this work is competitive both in
accuracy and in computational efficiency when compared with a standard
method for inverting banded matrices. This holds true both with and
without pivoting. We stress we are comparing two algorithms which are not yet thoroughly 
optimized (as LAPACK is).


\section{Concluding remarks}
\label{conclusions}

In this paper, we have introduced a new linearly scaling method to invert the banded matrices
that so often appear in problems of Computational Physics, Chemistry and
other disciplines. We have proven that this new algorithm is capable of
being more accurate than standard methods based on Gaussian elimination
at a lower computational cost, which opens the door to its use in many
practical problems, such as the ones described in the introduction.

Moreover, we have produced the analytical expressions that allow us to directly
obtain, in a recursive manner, the solution to the associated linear system in
terms of the entries of the original matrix. To have these
explicit formulae (which have also been presented for the calculation of the
full inverse matrix in the Appendix) at our disposal not only simplifies the task of coding
the needed computer algorithms, but it may also be useful to facilitate
analytical developments in the problems in which banded matrices appear.

In addition, we have checked its performance for general trial systems, and proven its 
usefulness for real physical problems (calculations on dynamics of proteins).

\section*{Aknowledgments}

The authors would like to thank J. L. Alonso, G. Ciccotti, J. M. Pe\~na, S.R. Christensen and
\'A. Rubio for illuminating discussions and useful advice, and M.
Garc\'ia-Risue\~no for helping with the plots, as well as the staff of
Caesaraugusta supercomputing facility (RES), where the test calculations of
this paper were run. This work has been supported by the research projects
E24/3 (DGA, Spain), FIS2009-13364-C02-01 (MICINN, Spain) 200980I064 (CSIC,
Spain) and ARAID and Ibercaja grant for young researchers (Spain). P. G.-R. is
supported by a JAE PREDOC grant (CSIC, Spain).

\appendix 

\section{Inverse of a banded matrix}
\label{sec_inversion}

In the previous sections, we proved that the banded linear system of $n$
equations with $n$ unknowns in~(\ref{basic_system}) can be solved in order $n$
operations. Sometimes, we are interested in obtaining the inverse matrix
$A^{-1}$ itself. We can do this in order $n^{2}$ operations using the same kind
of ideas discussed in the main body of the article. It should be stressed
that the explicit inverse of an arbitrary banded matrix usually cannot be
obtained in $\mathcal{O}(n)$ floating point operations, since the inverse of a banded matrix has
$n^{2}$ entries and it is not, in general, a banded matrix itself (an
exception to this is a block diagonal matrix). In order to obtain an efficient
way to invert $A$, we will derive some recursive relations
between the rows of $P$ (and $Q$). To this end, we will first calculate the
explicit expression of the entries of these matrices.

Using eqs. (\ref{PQcompletas}), (\ref{defP}), and (\ref{defQ}), from which
(\ref{GP_a}, \ref{GP_b}, \ref{GP_c}) and (\ref{6MB}) follow, and after some straightforward but long
calculations, one can show that the aforementioned entries satisfy
\numparts
\begin{eqnarray}\label{ssp}
P_{IJ} & = \xi_{JJ} \sum_{c \in C_{I,J,m_u}^\uparrow}
 \prod_{(K,L) \in c} \xi_{KL} & \textrm{for $I<J$} \ , \\
P_{II} & = \xi_{II} \ , & \\
P_{IJ} & = 0 & \textrm{for $I>J$} \ ,
\end{eqnarray}
\endnumparts
and
\numparts
\begin{eqnarray}\label{ssq}
Q_{IJ} & = 	\sum_{c \in C_{I,J,m_l}^\downarrow}
	 \prod_{(K,L) \in c} \xi_{KL} & \textrm{for $I>J$} \ , \\
Q_{II} & = 1 \ , & \\
Q_{IJ} & = 0 & \textrm{for $I<J$} \ .
\end{eqnarray}
\endnumparts
We call the summations appearing in the first line of each of these groups of
expressions \emph{jump summations}. There are two jump summations here, one
from $I$ to $J$ with increasing indices
($\uparrow$) and $m_{u}$ neighbours, and a \emph{jump summation} from $I$ to $J$
with decreasing indices ($\downarrow$) and $m_{l}$ neighbours, respectively.
The jump summation provides us with an explicit expression for all entries in
$A^{-1}$, without the need to recursively refer to other entries. This can be
useful in order to parallelize its calculation.

As it can be seen in the expressions, that each product in the sums contains a
number of coefficients $\xi_{KL}$. The pairs of indices $(K,L)$ which are
included in a given product are taken from a set $c$. In turn, each term of
the sum corresponds to a different set of pairs of indices $c$ drawn from a set
of sets of pairs of indices $C_{I,J,m_u}^\uparrow$ (in the case of $P_{IJ}$)
or $C_{I,J,m_l}^\downarrow$ (in the case of $Q_{IJ}$). Therefore, the only
detail that remains to understand these `jump summations' is to specify which
are the elements of these latter sets.

A given element $c$ of either $C_{I,J,m_u}^\uparrow$ or
$C_{I,J,m_l}^\downarrow$ can be expressed as
\begin{equation}
c = \{ (K_1,L_1), (K_2,L_2), \ldots, (K_S,L_S) \} \ ,
\end{equation}
in such a way that $C_{I,J,m_u}^\uparrow$ comprises all possible $c$'s 
that comply with a number of rules:
\begin{itemize}
\item $K_1 = I$, and $L_S = J$.
\item $K_r < L_r$, for $r=1,\ldots,S$.
\item $K_{r+1} = L_r$, for $r=1,\ldots,S$.
\item $L_r - K_r \leq m_u$, for $r=1,\ldots,S$.
\end{itemize}

Let us see an example:
\begin{eqnarray}
\sum_{c \in C_{3,6,2}^\uparrow} \prod_{(K,L) \in c} \xi_{KL}
 & = & \xi_{3,4} \xi_{4,5}\xi_{5,6} 
     + \xi_{3,4}  \xi_{4,6} + \xi_{3,5}  \xi_{5,6}  \ .
\end{eqnarray}

The rules to determine the elements of $C_{I,J,m_l}^\downarrow$ are
analogous to the ones above but they take into account that the indices
decrease:
\begin{itemize}
\item $K_1 = I$, and $L_S = J$.
\item $K_r > L_r$, for $r=1,\ldots,S$.
\item $K_{r+1} = L_r$, for $r=1,\ldots,S$.
\item $K_r - L_r \leq m_l$, for $r=1,\ldots,S$.
\end{itemize}

An example would be:
\begin{eqnarray}
\sum_{c \in C_{5,1,3}^\downarrow} \prod_{(K,L) \in c} \xi_{KL}
 & = & \xi_{5,4} \xi_{4,3} \xi_{3,2} \xi_{2,1}
     + \xi_{5,3} \xi_{3,2} \xi_{2,1}
     + \xi_{5,4} \xi_{4,2} \xi_{2,1}
 \nonumber \\
 &  & \mbox{} + \xi_{5,4} \xi_{4,3} \xi_{3,1}
              + \xi_{5,3} \xi_{3,1}
              + \xi_{5,2} \xi_{2,1}
              + \xi_{5,4} \xi_{4,1} \ .
\end{eqnarray}

If we first focus on $P$, it is easy to see that, according to the properties
of the jump summation, we have
\begin{eqnarray}
\label{jsum_recursive}
\sum_{c \in C_{I,J,m_u}^\uparrow} \prod_{(K,L) \in c} \xi_{KL} & = &
 \xi_{I,I+1} \sum_{c \in C_{I+1,J,m_u}^\uparrow}
 \prod_{(K,L) \in c} \xi_{KL}  \nonumber \\
 & & \mbox{} + \xi_{I,I+2} \sum_{c \in C_{I+2,J,m_u}^\uparrow}
		       \prod_{(K,L) \in c} \xi_{KL}  \nonumber \\
 & & \mbox{} + \ldots + \xi_{I,I+m_u} \sum_{c \in C_{I+m_u,J,m_u}^\uparrow}
                        \prod_{(K,L) \in c} \xi_{KL} \ .
\end{eqnarray}

If we insert a multiplicative factor $\xi_{JJ}$ at both sides and use
(\ref{ssp}), this equation becomes equation~(\ref{4MB}) obtained in
sec.~\ref{sec_main}:
\begin{equation}
P_{IJ}=\sum_{K=I+1}^{\min\{I+m_{u},n\}}\xi_{IK}P_{KJ} \qquad
  \textrm{for } I<J \ . \nonumber
\end{equation}

An analogous expression for $Q$ can be obtained in a similar way:
\begin{equation} 
Q_{IJ}=\sum_{L=J+1}^{\min\{J+m_{l},n\}}Q_{IL}\xi_{LJ}
 \qquad \textrm{for }I>J \ . \nonumber
\end{equation}

Now, if we define $\mu_{1}:=\min\{m_{u},J-I\}$, and
$\mu_{2}:=\min\{m_{l},I-J\}$, since $A^{-1}=PQ$ (see (\ref{basis})), we have
that
\begin{eqnarray}
\label{invA1}
(A^{-1})_{IJ} & = & (PQ)_{IJ} = \sum_{K=1}^{n}P_{IK}Q_{KJ} 
 = \sum_{K=J}^{n}P_{IK}Q_{KJ}= \nonumber \\
 & = & \sum_{K=J}^{n} \left(\sum_{L=I+1}^{I+\mu_{1}} P_{LK}\xi_{IL} 
  \right)Q_{KJ} = \sum_{L=I+1}^{I+\mu_{1}}\xi_{IL}
  \left(\sum_{K=J}^{n}P_{LK}Q_{KJ}\right) \nonumber \\
 & = & \sum_{L=I+1}^{I+\mu_{1}}\xi_{IL}(A^{-1})_{LJ} 
       \qquad \textrm{for $I<J$} \ ,
\end{eqnarray}
which is a recursive relationship for the superdiagonal entries of $A^{-1}$. 

Performing similar computations, we have
\begin{eqnarray}
(A^{-1})_{II} & = & \xi_{II} + \sum_{L=I+1}^{I+\mu_{1}}\xi_{IL}(A^{-1})_{LJ}
 = \xi_{II} + \sum_{L=J+1}^{J+\mu_{2}}\xi_{LJ}(A^{-1})_{IL}  \ , 
  \label{invA2} \\
(A^{-1})_{IJ} & = & \sum_{L=J+1}^{J+\mu_{2}}\xi_{LJ}(A^{-1})_{IL} \qquad 
 \textrm{for $I>J$} \ . \label{invA3}
\end{eqnarray}

Using the last three equations, we can easily construct an algorithm to
compute $A^{-1}$ in $\mathcal{O}(n^{2})$ floating point operations. This algorithm would first
calculate $(A^{-1})_{nn}=\xi_{nn}$. Then it would use (\ref{invA1}) to obtain,
in this order, $(A^{-1})_{n-1,n}$, $(A^{-1})_{n-2,n}$, $\ldots$,
$(A^{-1})_{1n}$. These are the superdiagonal ($I<J$) entries of the $n$-th
column. Then, it would use (\ref{invA3}) to obtain, in this order,
$(A^{-1})_{n,n-1}$, $(A^{-1})_{n,n-2}$, $\ldots$, $(A^{-1})_{n1}$, i.e., the
subdiagonal ($I>J$) entries of the $n$-th row. Once the $n$ row and column of
$A^{-1}$ are known, $(A^{-1})_{n-1,n-1}$ can be obtained with (\ref{invA2}).
Then (\ref{invA1}) and (\ref{invA3}) can be used to obtain the entries of
this ($n-1$) column and row, respectively. When calculating the entries of a
column $J$, i.e., $\xi_{KJ}$ with $K<J$, $\xi_{LJ}$ is always obtained
before $\xi_{L-1,J}$. When calculating the entries of a row $I$, i.e.,
$\xi_{IK}$ with $I>K$, $\xi_{IK}$ is always obtained before
$\xi_{I,K-1}$. This procedure can be repeated for all rows and columns of
$A^{-1}$, and the calculation of the $K$-th row and column can be performed in
parallel.



\section*{References}
\bibliography{refs}

\providecommand{\newblock}{}
\begin{thebibliography}{10}
\expandafter\ifx\csname url\endcsname\relax
  \def\url#1{{\tt #1}}\fi
\expandafter\ifx\csname urlprefix\endcsname\relax\def\urlprefix{URL }\fi
\providecommand{\eprint}[2][]{\url{#2}}

\bibitem{Don1987PC}
Dongarra J and Johnson S~L 1987 {\em Parallel Computing\/} {\bf 5} 219--246

\bibitem{Hym2002CG}
Hyman J, Morel J, Shashkov M and Steinberg S 2002 {\em Computational
  Geosciences\/} {\bf 6} 333--352

\bibitem{Sha1997IJCM}
Shaw R~E and Garey L~E 1997 {\em International Journal of Computer
  Mathematics\/} {\bf 65, 1-2} 121--129

\bibitem{pap1991PC}
Paprzycki M and Gladwell I 1991 {\em Parallel Computing\/} {\bf 17} 133--153

\bibitem{Wri1992SIAM}
Wright S~J 1992 {\em SIAM J. Sci. Stat. Comput.\/} {\bf 13} 742--764

\bibitem{Bri1977JCOP}
Briley W~R and McDonald H 1977 {\em JCOP\/} {\bf 24, 4} 372--397

\bibitem{Ari1992AM}
Ariel P~D 1992 {\em Acta Mechanica\/} {\bf 103} 31--43

\bibitem{Had2008}
Haddad O~M, Al-Nimr M~A and Shatnawi G~H 2008 {\em Selected Papers from the
  WSEAS Conferences in Spain, September 2008 Santander, Cantabria, Spain\/}

\bibitem{Had2004E}
Haddad O, Abuzaid M and Al-Nimr M 2004 {\em Entropy\/} {\bf 6, (5)} 413--416

\bibitem{Pol2006PC}
Polizzi E and Shameh A~H 2006 {\em Parallel Computing\/} {\bf 32} 177--194

\bibitem{Pol2004JCOP}
Polizzi E and Ben~Abdallah N 2004 {\em Journal of Computational Physics\/} {\bf
  202, 1} 150--180

\bibitem{Lum1988}
Lumsdaine A, White J, Webber D and Sangiovanni-Vincentelli A 1988 {\em Research
  Laboratory of Electronics Dept. of Electrical Engineering and Computer
  Science Massachusetts Institute of Technology Cambridge, MA 02139,
  CH2657-5/88/0000/0308 01.000 1988IEEE\/}

\bibitem{San1986JCOP}
Sanz-Serna J~M and Christie I 1986 {\em Journal of Computational Physics\/}
  {\bf 67, 2} 348--360

\bibitem{Gua1999JCP}
Guantes R and Farantos S~C 1999 {\em Journal of Chemical Physics\/} {\bf 111,
  24} 10827--10835

\bibitem{Gua1999JCOP}
Guardiola R and Ros J 1999 {\em Journal of Computational Physics\/} {\bf 111,
  24} 374--389

\bibitem{Cas2006PSS}
Castro A, Appel H, Oliveira M, Rozzi C~A, Andrade X, Lorenzen F, Marques M~A~L,
  Gross E~K~U and Rubio A 2006 {\em Phys. Stat. Sol\/} {\bf 243} 2465

\bibitem{Mar2003CPC}
Marques M~A~L, Castro A, Bertsch G~F and Rubio A 2003 {\em Comp. Phys. Comm.\/}
  {\bf 151} 60

\bibitem{Ryc1977JCOP}
Ryckaert J~P, Ciccotti G and Berendsen H~J~C 1977 {\em J. Comput. Phys.\/} {\bf
  23} 327--341

\bibitem{Alv2004JPCM}
Alvarez-Estrada R~F and Calvo G~F 2004 {\em Journal of Physics: Condensed
  Matter\/} {\bf 16} S2037

\bibitem{Alv2005JPCM}
Calvo G~F and Alvarez-Estrada R~F 2005 {\em Journal of Physics: Condensed
  Matter\/} {\bf 17} 7755

\bibitem{Maz2007JPA}
Mazars M 2007 {\em J. Phys. A: Math. Theor.\/} {\bf 40, 8} 1747--1755

\bibitem{GR2011JCC}
Garc\'ia-Risue\~no P, Echenique P and Alonso J~L 2011 {\em J. Comput. Chem.\/}
  {\bf 32} 3039--3046

\bibitem{Str1969NM}
Strassen V 1969 {\em Numerische Mathematik\/} {\bf 13} 354–--356

\bibitem{Alo2008PRL}
Alonso J~L, Andrade X, Echenique P, Falceto F, Prada-Gracia D and Rubio A 2008
  {\em Phys. Rev. Lett.\/} {\bf 101} 096403

\bibitem{Has1970bio}
Hastings W~K 1970 {\em Biometrika\/} {\bf 57, 1} 97--109

\bibitem{Ech2007MP}
Echenique P and Alonso J~L 2007 {\em Mol. Phys.\/} {\bf 105} 3057--3098

\bibitem{Gri1993PRA}
Gritsenko O~V, Rubio A, Balbás L~C and Alonso J~A 1993 {\em Phys. Rev. A\/}
  {\bf 47} 1811--1816

\bibitem{Pea1995CoPCo}
Pearlman D~A, Case D~A, Caldwell J~W, Ross W~R, Cheatham~III T~E, DeBolt S,
  Ferguson D, Seibel G and Kollman P 1995 {\em Comp. Phys. Commun.\/} {\bf 91}
  1--41

\bibitem{claudioFF1}
Cavasotto C~N and W~Orry A~J May {\em Current Topics in Medicinal Chemistry\/}
  {\bf 7} 1006--1014

\bibitem{claudioFF2}
Anisimov V~M and Cavasotto C~N 2011 {\em Journal of Computational Chemistry\/}
  {\bf 32} 2254--2263

\bibitem{dft_linear_scaling_1}
Hine N, Haynes P, Mostofi A, Skylaris C~K and Payne M 2009 {\em Computer
  Physics Communications\/} {\bf 180} 1041 -- 1053

\bibitem{siesta2002}
Soler J~M, Artacho E, Gale J~D, Garc\'ia A, Junquera J, Ordej\'on P and
  S\'anchez-Portal D 2002 {\em Journal of Physics: Condensed Matter\/} {\bf 14}
  2745

\bibitem{conquest_linear_dft}
Gillan M, Bowler D, Torralba A and Miyazaki T 2007 {\em Computer Physics
  Communications\/} {\bf 177} 14 -- 18 proceedings of the Conference on
  Computational Physics 2006 - CCP 2006, Conference on Computational Physics
  2006

\bibitem{And1999BOOK}
Anderson E, Bai Z and Bischof C~e~a (1999) {\em LAPACK User's Guide\/} release
  3.0 ed (Philadelphia: SIAM)

\bibitem{Pre2007BOOK}
Press W~H, Teukolsky S~A, Vetterling W~T and Flannery B~P (2007) {\em Numerical
  recipes. {T}he art of scientific computing\/} 3rd ed (New York: Cambridge
  University Press)

\bibitem{WatXXXXBOOK}
Wakins D~S 1991 {\em Fundamentals of Matrix Computations\/} 2nd ed (New York:
  {W}iley {I}nter-{S}cience)

\bibitem{Gol1989BOOK}
Golub G~H and Van~Loan C~F (eds) 1993 {\em Matrix Computations\/} 2nd ed
  (Baltimore and London: The Johns Hopkins University Press)

\bibitem{Cas2004JCP}
Castro A, Marques M~A~L and Rubio A 2004 {\em J. Chem. Phys.\/} {\bf 121}
  3425--3433

\bibitem{Cha2003SIAM}
Chandrasekaran S and Gu M 2003 {\em SIAM Journal on Matrix Analysis and its
  Applications\/} {\bf 25(2)} 373--384

\bibitem{Bin1999SIAM}
Bini D~A and Meini B 1999 {\em SIAM J. Matrix Anal. Appl.\/} {\bf 20}
  700–--719

\bibitem{Mei1985PC}
Meier U 1985 {\em Parallel Computing\/} {\bf 2} 33--23

\bibitem{Zha1994PC}
Zhang H and Moss W~F 1994 {\em Parallel Computing\/} {\bf 20, 8} 1089--1105

\bibitem{Law1984ACM}
Lawrie D~H and Sameh A~H 1984 {\em ACM Transactions on Mathematical Software\/}
  {\bf 10, 2} 185--195

\bibitem{Joh1985ACM}
Johnson S~L 1985 {\em ACM Transactions on Mathematical Software\/} {\bf 11}
  271--288

\bibitem{Che1978ACM}
Chen S~C, Kuck D~J and Sameh A~H 1978 {\em ACM Transactions on Mathematical
  Software\/} {\bf 4} 270--277

\bibitem{Eva1976CJ}
Evans D~J and Hatzopoulos M 1976 {\em The Computer Journal\/} {\bf 19, 2}
  184--187

\bibitem{Gar2000AMC}
Garey L~E and Shaw R~E 2000 {\em Applied Mathematics and Computation\/} {\bf 5,
  311} 133--143

\bibitem{Arb1994TR}
Arbenz P and Gander W 1994 {\em Technical Report TR 221, Inst. for Scientific
  Comp., ETH, Z\"urich\/}

\bibitem{Gol2001NLAA}
Golub G~H, Sameh A~H and Sarin V 2001 {\em Numerical linear algebra with
  applications\/} {\bf 8} 297--316

\bibitem{Don1984PC}
Dongarra J and Sameh A~H 1984 {\em Parallel Computing\/} {\bf 1} 223--235

\bibitem{HenXXXXBOOK}
Hennessy J~L and Patterson D~A 2003 {\em Computer Architecture, A Quantitative
  Approach\/} 3rd ed (San Mateo, CA: Morgan Kaufmann - Elsevier)

\bibitem{libro_HPC}
Hager G and Wellein G 2011 {\em Introduction to High Performance Computing for
  Scientists and Engineers\/} 1st ed (CRC Press - Taylor \& Francis Group)

\bibitem{Leimkuhlerbook}
Leimkuhler B and Reich S 2004 {\em Simulating Hamiltonian dynamics\/} 1st ed
  (Cambridge University Press - Cambridge monographs on applied and
  computational Mathematics)

\bibitem{Hes1997JCC}
Hess B, Bekker H, Berendsen H~J~C and Fraaije J~G~E~M 1997 {\em J. Comput.
  Chem.\/} {\bf 18} 1463--1472

\bibitem{Avogadro}
 2010 Avogadro: an open-source molecular builder and visualization tool.
  version 1.0.1 \url{http://avogadro.openmolecules.net/}

\end{thebibliography}

\end{document}